\documentclass[aps,prd,preprint,groupedaddress,showpacs]{revtex4-2}
\usepackage{subcaption}

\usepackage{epsfig}
\usepackage{graphics}
\usepackage{slashed}
\usepackage{color}
\usepackage[citecolor=true]{hyperref}
\usepackage{multirow}
\usepackage{amsmath}

\begin{document}

\leftmargin -2cm
\def\choosen{\atopwithdelims..}

\boldmath
\title{Single isolated photon production in the NLO${}^\star$ \\
approximation of the Parton Reggeization Approach} \unboldmath

\author{\firstname{A.A.}~\surname{Chernyshev}} \email{chernyshev@theor.jinr.ru}

\affiliation{Moscow State U., Dubna Branch, Leningradskaya st., 12,
141980, Dubna, Russia}

\affiliation{Joint Institute for Nuclear Research, 141980, Dubna, Russia}

\author{\firstname{V.A.}~\surname{Saleev}} \email{saleev.vladimir@gmail.com}

\affiliation{Samara National Research U., Moskovskoe Shosse,
34, 443086, Samara, Russia}

\affiliation{Joint Institute for Nuclear Research, 141980, Dubna, Russia}

\begin{abstract}
The hadroproduction of single isolated photon at high energies is
studied within the framework of the NLO${}^\star$ approximation of
the Parton Reggeization Approach based on the modified Multi--Regge
limit of the hard scattering QCD amplitudes. The contribution from
the LO subprocess $Q \bar Q \to \gamma$ and the NLO tree--level
corrections $Q R \to \gamma q$ and $Q \bar Q \to \gamma g$ are
considered. To avoid specific double counting in rapidity between
tree--level corrections to the hard scattering coefficient and
unintegrated PDFs, a corresponding subtraction scheme is proposed.
We demonstrated self--consistency of the Parton Reggeization
Approach using the subtraction scheme. The results of calculations
are compared with data from various Collaborations in the wide
energy range $\sqrt{S} = 24 \ {\rm GeV} - 13 \ {\rm TeV}$. We
obtained a quite satisfactory agreement with data up to $p_T^\gamma
/ \sqrt{S} \simeq 0.2 - 0.3$. Predictions for the future SPD NICA
experiment were discussed.

\end{abstract}


\maketitle

\section{Introduction}\label{sec:intro}

The process of direct photon hadroproduction is the subject of the
intensive
experimental~\cite{D0:2005,ATLAS:2016,ATLAS:2019,ATLAS:2023,CMS:2018,
PHENIX:2012,UA6:1998} and
theoretical~\cite{Gordon:1994,Catani:2002,Aurenche:2006,Saleev:2008,Hameren:2023}
studies especially at the LHC energies $\sqrt{S} = 7 - 13$ TeV
because of several aspects. Firstly, since this process is a {\em
single--scale} $\mu_F \sim p_T^\gamma$, where $\mu_F$ is a
factorization scale of the process and $p_T^\gamma \equiv |{\bf
p}_T^\gamma|$ is a photon transverse momentum, they considered as an
important probe of the perturbative QCD in the fixed order Collinear
Parton Model (CPM) calculations at high $p_T^\gamma$ where the
higher twist corrections are strongly power suppressed ${\cal
O}\left( \Lambda^\# / \mu_F^\# \right)$. Additionally, this process
is a good tool to study gluon parton distribution functions (PDFs)
at the LHC, RICH and in the future EIC and SPD NICA
experiments~\cite{SPD}. The CPM includes only the
Dokshitzer--Gribov--Lipatov--Altarelli--Parisi (DGLAP) resummation
of large logs $\alpha_S^n \, \ln^n \mu_F^2 / \Lambda^2$ into the
PDFs~\cite{GL:1972,AP:1977,D:1977}. The impact of the leading order
(LO) CPM contributions is not enough to describe data, moreover,
complete next--to--leading order (NLO) predictions steel
underestimate data by about $10 - 20 \%$ in some kinematical region,
for example, see parton level Monte--Carlo (MC) generator {\tt
JetPhoX}~\cite{Catani:2002} NLO predictions in
Refs.~\cite{ATLAS:2016,ATLAS:2019,CMS:2018,ATLAS:2023}.

At high energies, instead of fixed order CPM calculations one may
use the {\em High--Energy Factorization} (HEF) approach also known
as {\em $k_T$--factorization} which was formulated in
Refs.~\cite{HEF1,HEF2,HEF3}. The main idea of the HEF is to resum
already in the LO approximation of the CPM large radiative
corrections enhanced by the logs $\alpha_S^n \ln^n \hat s / (- \hat
t)$ with $\hat s \gg (- \hat t)$ into the {\em unintegrated parton
distribution functions} (uPDFs). For the first time such resummation
was done in the Balitsky--Fadin--Kuraev--Lipatov (BFKL)
equation~\cite{BFKL1,BFKL2,BFKL3,BFKL4}, which based on the {\em
Multi--Regge kinematics} (MRK) limit~\cite{BFKL4} of the QCD
high--energy scattering amplitudes. The BFKL approach valid only in
the Regge limit of the QCD where the following hierarchy of the
"$\pm$"--components of the $4$--momenta between particles produced
in the $\hat t$--channel chain is imposed: $q^\mp \ll |{\bf q}_T|
\sim \mu_F \sim q^\pm \ll \sqrt{S}$, for the notations see
Sec.~\ref{sec:PRA}. In the paper, we use the {\em Parton
Reggeization Approach} (PRA)~\cite{PRA1,PRA2,PRA3,NS:2020} which is
a gauge--invariant formulation of the HEF approach. The PRA based on
the {\em modified MRK} (mMRK) approximation of the hard QCD
amplitudes~\cite{NS:2016,PRA3,NS:2020}, which includes Reggeized
amplitudes constructed according to the Feynman rules of the
Effective Field Theory (EFT) for processes in the MRK proposed by
L.N.~Lipatov~\cite{Lipatov:1,Lipatov:2,Lipatov:3}, and
Kimber--Martin--Ryskin--Watt (KMRW) model~\cite{KMR,MRW} of the
uPDFs, which was modified in Ref.~\cite{NS:2020}. The PRA accurate
correct both in collinear and MRK limits~\cite{PRA3}. The
consistency with the Collins--Soper--Sterman (CSS)
approach~\cite{CSS} valid in the region of small transverse momenta
${\bf q}_T^2 \ll \mu_F^2$ was demonstrated in Ref.~\cite{NS:2020}.
However, since the CSS formalism is suitable only for the low
transverse momenta of the final state particles
production~\cite{Collins:2011}, the HEF is the only way to study
effects of non--collinear parton dynamics in the single direct
photon production.

The PRA was already used for the description of various processes involving
direct photons. Production of the single direct photon first was studied in
Ref.~\cite{Saleev:2008} within the LO approximation of the PRA
at the Tevatron energy $\sqrt{S} = 1.8$ TeV.
In Ref.~\cite{NS:2014}, the study of the isolated photon plus jet
photoproduction was done in the PRA.
The diphoton production was studied in the PRA twice: first at the Tevatron
energies in Ref.~\cite{Saleev:2009} with the impact of the LO subprocess,
and the second one in Ref.~\cite{NS:2016} at Tevatron and LHC energies,
where the tree--level NLO (NLO${}^\star$) corrections of the order
${\cal O}\left( \alpha^2 \alpha_S^1 \right)$ and quark
box contribution ${\cal O}\left( \alpha^2 \alpha_S^2 \right)$
were added. Production of the three isolated photons was
studied in the PRA with the help of the {\tt KaTiE}~\cite{katie} event generator in
Ref.~\cite{KS:2022}. In the two last processes the impact of the NLO${}^\star$
corrections like quark--gluon scattering is found to be important, so it is
interesting to study the single direct photon production within the NLO${}^\star$
approximation of the PRA once again using the new uPDFs proposed in
Ref.~\cite{NS:2020}.

The paper has the following structure. In the Sec.~\ref{sec:PRA}, we
discuss the main aspects of the PRA, which are important for the
discussion of the tree--level corrections calculations in Sec.~\ref{sec:phot}.
The results of our calculations are presented in Sec.~\ref{sec:res}, as
well as comparison with the experimental data at the wide energy range.
Our conclusions are collected in Sec.~\ref{sec:conc}.

\section{Parton Reggeization approach}\label{sec:PRA}

In the study, we use Sudakov decomposition of $4$--momenta without
separating covariant and contravariant "$\pm$" indices using
light--cone basis vectors $n_\pm$ with normalization $(n_\pm, n_\mp)
= 2$ in the form: $p = p_L + p_T$, where $p_L = \left( p^+ n_- + p^-
n_+ \right) / 2$ and $p_T$ are the longitudinal and transverse
components respectively. We choose $n_\pm = ( 2 / \sqrt{S} ) \,
P_{1, 2}$, where $P_{1, 2}$ are the proton's momenta and $S = 2
\left( P_1, P_2 \right)$. A "$\pm$" components obtained as a
projections $p^\pm = (p, n_\pm)$. In this notation, rapidity defined
as $y(p) = (1 / 2 ) \, \ln \left( p^+ / p^- \right)$ and $p^2 = p^+
p^- - {\bf p}_T^2$. We also use notation $\slashed{p} = p_\mu
\gamma^\mu$.

\begin{figure}[ht]
\centering
\includegraphics[scale=.8]{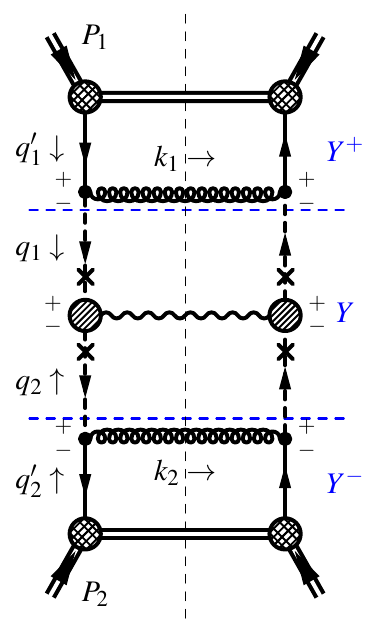}
\caption{Schematic representation of the mMRK amplitude.}\label{fig:1}
\end{figure}

To derive the PRA factorization formula for the direct photon
production, we may consider auxiliary $t$--channel scattering
subprocess for the direct photon production in the collinear quarks
$q_{1,2}'^2 \simeq 0$ annihilation with the emission of two
additional on mass--shell gluons $k_{1,2}^2 = 0$, for a diagramming
representation also see Fig.~\ref{fig:1}:
\begin{eqnarray}
q \ (q_1') + \bar q \ (q_2') & \to & g \ (k_1) + \gamma \ (q_3) + g \ (k_2),
\label{pr:1}
\end{eqnarray}
$t$--channel momenta are $q_{1,2} = q_{1,2}' - k_{1,2}$. For the
further discussion, it is suitable to introduce dimensionless
momenta fractions of the large light--cone components as $z_+ =
q_1^+ / q_1'^+$ and $z_- = q_2^- / q_2'^-$. In the MRK limit, the
strong hierarchy of the "$\pm$"--components is imposed: $q_1'^+ \gg
q_1^+$ and $q_2'^- \gg  q_2^-$. In this kinematical regime,
$t$--channel partons with momenta $q_{1,2}$ are Reggeized, which
leads to the special form of the scattering amplitudes, see
Ref.~\cite{Lipatov:1996} for an overview. The interactions of the
standard QCD degrees of freedom with the Reggeized ones are
described by {\em effective vertices}. To obtain effective vertices
one may use Lipatov's EFT formalism proposed for Reggeized gluons in
Ref.~\cite{Lipatov:1} and for Reggeized quarks in
Ref.~\cite{Lipatov:2}. Feynman rules of the EFT was constructed in
Ref.~\cite{Lipatov:3}.

\begin{figure}[ht]
\centering
\includegraphics[scale=0.7]{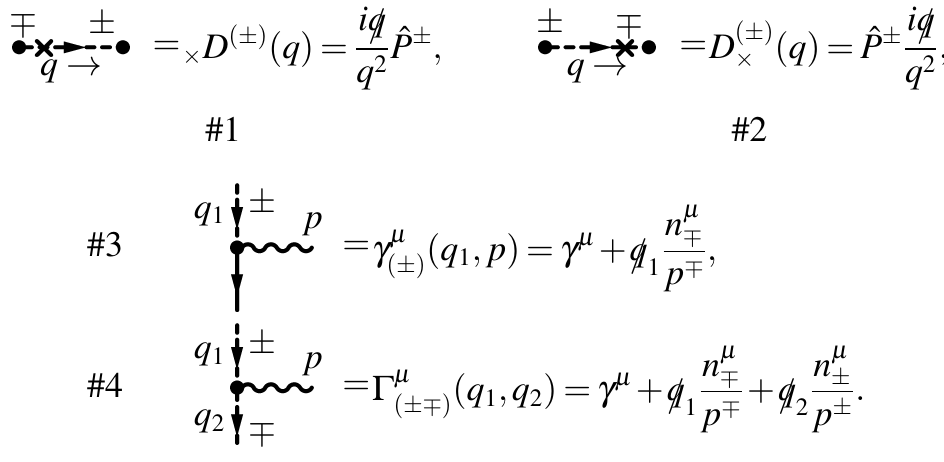}
\caption{Feynman rules of the
EFT~\cite{Lipatov:1,Lipatov:2,Lipatov:3}: $\# 1$ and $\# 2$ -- mMRK
Reggeized quark propagators~\cite{NS:2016}, $\# 3$ -- $Q \bar q
\gamma$ effective vertex, and $\# 4$ -- Fadin--Sherman $Q \bar Q
\gamma$ effective vertex~\cite{Fadin:1977}. All momenta in vertices
are incoming.} \label{fig:2}
\end{figure}

Relevant for the derivation of the hard amplitudes in the mMRK form,
Feynman rules of the EFT~\cite{Lipatov:1,Lipatov:2,Lipatov:3} for the readers
convenience are collected in Fig.~\ref{fig:2},
$\hat P^\pm = (1 / 4) \, \slashed{n}^\mp \slashed{n}^\pm$ is a projector
operator. All other relevant for the study effective vertices may be obtained
with the help of the {\tt ReggeQCD} model--file by M. Nefedov for
{\tt FeynArts}~\cite{FeynArts}.
Using this Feynman rules, squared matrix element (ME) for the
subprocess~(\ref{pr:1}) reads:
\begin{align}
\overline{\mid {\cal M}^{\rm (mMRK)} (q \bar q \to g \gamma g) \mid^2} & =
\left( \frac{g^2 C_F}{2 N_c} \right)^2
\nonumber \\ \times
{\rm tr} \bigg[
\theta & \left( y(q_3) - y(k_2) \right)
\hat P^-
\frac{\slashed{q}_2}{q_2^2}
\gamma_\mu^{(+)}(q_2, k_2) \,
\slashed{q}'_2 \, \gamma^\mu_{(+)}(q_2, k_2) \, \frac{\slashed{q}_2}{q_2^2} \,
\hat P^+
{\cal A}_\gamma^{\rm (MRK)\dagger}
\nonumber \\ \times
\theta & \left( y(k_1) - y(q_3) \right)
\hat P^+ \,
\frac{\slashed{q}_1}{q_1^2}
\gamma_{(-)}^\nu(q_1, k_1) \,
\slashed{q}'_1 \, \gamma^{(-)}_\nu(q_1, k_1) \, \frac{\slashed{q}_1}{q_1^2} \,
\hat P^-
{\cal A}_\gamma^{\rm (MRK)}
\bigg], \label{eq1}
\end{align}
here $C_F = 4 / 3$ is a Casimir operator value for the ${\rm
SU}(N_c)$ group with $N_c = 3$ colors, rapidity ordering
$\theta$--functions applied by the Reggeized quark
propagator~\cite{Lipatov:2} omitted in the Feynman rules in
Fig.~\ref{fig:2}. Color and Lorentz indices of the amplitude ${\cal
A}_\gamma^{\rm (MRK)} \equiv {\cal A}^{\rm (MRK)}(Q \bar Q \to
\gamma)$ are omitted, $Q (\bar Q)$ denote Reggeized quark
(antiquark). In the LO approximation, the amplitude ${\cal
A}_\gamma^{\rm (MRK)}$ is expressed through the well known
Fadin--Sherman effective vertex~\cite{Fadin:1977} obeys
Slavnov--Teylor identity: $(q_1 + q_2)_\mu \Gamma^\mu_{(+-)}(q_1,
q_2) \equiv 0$, see Fig.~\ref{fig:2}.

The ME~(\ref{eq1}) may be factorized using the {\em mMRK
approximation}~\cite{NS:2016,PRA3,NS:2020}.
The main idea of the mMRK is to keep the exact kinematics in the region
up to a mark "$\times$" on the Reggeized quark propagator. The position
of the projector operator doesn't aspect when squaring the amplitude,
but important for the mMRK factorization, see discussion in Ref.~\cite{NS:2016}.
For the MRK hierarchy of the "$\pm$"--components between exchange partons to be
valid, the virtualities of
the Reggeized partons should arise mainly from the transverse components
of the momenta: $| q_1^+ \, q_1^- | \ll {\bf q}_{T_1}^2$ and
$| q_2^+ \, q_2^- | \ll {\bf q}_{T_2}^2$,
see Sec.~2 in Ref.~\cite{MRW}. Using this relation
and the momenta conservation law $q_{1,2} = q_{1,2}' - k_{1,2}$, one can obtain
important kinematical constraint on the virtualities of the Reggeized partons
in the chain:
${\bf q}_{T_{1,2}}^2 > \left( z_\pm / (1 - z_\pm) \right) {\bf k}_{T_{1,2}}^2$.
While $q_{1, 2}' = \left( q_{1,2}'^\pm / 2 \right) n_\mp$,
where $q_{1, 2}'^{\pm} = \tilde x_{\pm} \, P_{1, 2}^\pm$,
using the constraint, $4$--momenta may be written in the form:
\begin{eqnarray}
q_1 & = &
\frac{z_+ q_1'^+}{2} n_- - \frac{{\bf q}_{T_1}^2}{2 (1 - z_+) q_1'^+} n_+ +
q_{T_1}, \label{q1} \\
q_2 & = &
- \frac{{\bf q}_{T_2}^2}{2 (1 - z_-) q_2'^-} n_- + \frac{z_- q_2'^-}{2} n_+  +
q_{T_2}, \label{q2}
\end{eqnarray}
so that $q_{1,2}^2 = - {\bf q}_{T_{1,2}}^2 / (1 - z_\pm)$.
In the MRK limit $z_\pm \ll 1$ and the virtualities are not limited.

With the help of virtuality's constraints,
rapidity ordering $\theta$--functions may be rewritten using
KMRW--cutoff function~\cite{KMR}
$\Delta\left( {\bf q}_T^2, \mu^2 \right) =
\mu / \left( \mu + |{\bf q}_T| \right)$,
then using the mMRK approximation together with the exact form of the
momenta~(\ref{q1})--(\ref{q2}), the ME~(\ref{eq1}) may be factorized in the
following way\cite{PRA2,NS:2020}:
\begin{align}
\overline{\mid {\cal M}^{\rm (mMRK)} (q \bar q \to g \gamma g) \mid^2} =
\frac{4 \, g^2}{q_1^2 \, q_2^2} \
&
\frac{P_{q q}(z_+)}{z_+} \
\theta\left( \Delta\left( {\bf q}_{T_1}^2, \mu^2 \right) - z_+ \right)
\nonumber \\ \times &
\frac{P_{q q}(z_-)}{z_-} \
\theta\left( \Delta\left( {\bf q}_{T_2}^2, \mu^2 \right) - z_- \right)
\nonumber \\ \times &
\overline{\mid {\cal A}^{\rm (MRK)}
(Q ({\bf q}_{T_1}) \, \bar Q ({\bf q}_{T_2}) \to \gamma) \mid^2}
+ {\cal O}\left( \frac{\mu^2}{S} \right), \label{mMRK}
\end{align}
where
$P_{q q}(z) = C_F \left[ \left( 1 + z^2 \right) / \left( 1 - z \right) \right]_+$
is a usual DGLAP splitting function with "$+$"--prescription.
Here and below we put $\mu_F = \mu_R = \mu$.
Calculations of the PRA $2 \to 1$ ME
$\overline{\mid {\cal A}^{\rm (MRK)}(Q \bar Q \to \gamma) \mid^2}$
were first performed in Ref.~\cite{Saleev:2008}:
\begin{eqnarray}
\overline{\mid {\cal A}^{\rm (MRK)}
(Q ({\bf q}_{T_1}) \, \bar Q ({\bf q}_{T_2}) \to \gamma) \mid^2} & = &
4 \pi \alpha \ \frac{C_A}{N_c^2} \times
\left( {\bf q}_{T_1}^2 + {\bf q}_{T_2}^2 \right), \label{2To1}
\end{eqnarray}
here $C_A = 3$. In Ref.~\cite{PRA2}, the validity of the mMRK
factorization~(\ref{mMRK}) in the collinear ${\bf q}_{T_{1,2}}^2 \ll \mu^2$ with
arbitrary $z_{\pm} \in (0, 1)$ and MRK $z_\pm \to 1$ with arbitrary
${\bf q}_{T_{1,2}}^2$ limits has been proved.

The CPM cross section for the subprocess~(\ref{pr:1}) differential over the
phase--space of the photon
$d \Phi_\gamma = 1 / (2 \pi)^3 d^3 {\bf q}_3 / (2 q_3^0)$ is following:
\begin{eqnarray}
\frac{d\sigma^{\rm (CPM)}(p p \to g \gamma g X)}{d \Phi_\gamma} & = &
\int \frac{d\tilde x_+}{\tilde x_+} \ F_q(\tilde x_+, \mu^2)
\int \frac{d\tilde x_-}{\tilde x_-} \ F_{\bar q}(\tilde x_-, \mu^2)
\nonumber \\ & \times &
\int \frac{d k_1^+ d^2 {\bf k}_{T_1}}{(2 \pi)^3 \, 2 k_1^+} \
\int \frac{d k_2^- d^2 {\bf k}_{T_2}}{(2 \pi)^3 \, 2 k_2^-} \
\frac{(2 \pi)^4}{I(\tilde x_+, \tilde x_-)} \
\delta^{(4)}\left( q_1 + q_2 - q_3 \right)
\nonumber \\ & \times &
\overline{\mid {\cal M}(q \bar q \to g \gamma g) \mid^2} +
{\cal O}\left( \frac{\Lambda^\#}{\mu^\#} \right),
\end{eqnarray}
here $F(\tilde x_\pm, \mu^2) = \tilde x_\pm \, f(\tilde x_\pm,
\mu^2)$, $f(\tilde x_\pm, \mu^2)$ is a usual collinear PDF, and
$I(\tilde x_+, \tilde x_-) = 2 \tilde x_+ \tilde x_- S$ is a
flux--factor. Substituting the mMRK approximation~(\ref{mMRK}) of
the ME~(\ref{eq1}) and changing variables as it was done in
Ref.~\cite{PRA2}, one can obtain the $k_T$--factorization formula
for the direct photon production:
\begin{eqnarray}
\sigma^{\rm (LO, \ PRA)}(p p \to \gamma X) & = &
\int \frac{d x_+}{x_+} \int \frac{d^2 {\bf q}_{T_1}}{\pi} \
\Phi_q^{\rm (tree)}(x_+, {\bf q}_{T_1}^2, \mu^2)
\int \frac{d x_-}{x_-} \int \frac{d^2 {\bf q}_{T_2}}{\pi} \
\Phi_{\bar q}^{\rm (tree)}(x_-, {\bf q}_{T_2}^2, \mu^2)
\nonumber \\ & \times &
{\cal H}^{\rm (LO, \ mMRK)}_\gamma(x_\pm, {\bf q}_{T_{1,2}}, \mu^2)
+ {\cal O}\left( \frac{\Lambda^\#}{\mu^\#}, \frac{\mu^2}{S} \right), \label{PRA0}
\end{eqnarray}
here $x_\pm = \tilde x_\pm \, z_\pm$, hard scattering coefficient
${\cal H}^{\rm (LO, \ mMRK)}_\gamma(x_\pm, {\bf q}_{T_{1,2}}, \mu^2)$
is expressed in terms of the squared Reggeized MRK amplitude
$\overline{\mid {\cal A}^{\rm (MRK)}_\gamma \mid^2}$ in a standard way:
$$
\frac{d {\cal H}^{\rm (LO, \ mMRK)}_\gamma(x_\pm, {\bf q}_{T_{1,2}}, \mu^2)}
{d \Phi_\gamma} =
(2 \pi)^4 \ \delta^{(4)}\left( q_1 + q_2 - q_3 \right) \
\frac{\overline{\mid {\cal A}^{\rm (MRK)}
(Q ({\bf q}_{T_1}) \, \bar Q ({\bf q}_{T_2}) \to \gamma) \mid^2}}
{I(x_+, x_-)}.
$$

In the PRA, the quark uPDF is appeared in the so called
"tree--level" form, which is just a convolution of the DGLAP
splitting function $P_{i j}(z)$ with the PDF density $F_j(x / z,
\mu^2)$ multiplied by rapidity ordering $\theta$--function:
\begin{eqnarray}
\Phi_q^{\rm (tree)}(x, {\bf q}_T^2, \mu^2) & = &
\frac{\alpha_S(\mu^2)}{2 \pi} \frac{1}{{\bf q}_T^2}
\sum\limits_{j = g, q}
\int_x^1 dz \
P_{i j}(z) \ F_j\left( \frac{x}{z}, \mu^2 \right)
\theta\left( \Delta({\bf q}_T^2, \mu^2) - z \right), \label{uPDF1}
\end{eqnarray}
note that in the original KMRW model~\cite{KMR,MRW} rapidity ordering applied only on
gluons, while in the PRA it's imposed both on quarks and gluons.
Such defined uPDF has IR divergence ${\bf q}_T^2 \to 0$,
to resolve this problem, we require that modified KMRW (mKMRW) uPDF must
satisfy the exact normalization condition of the BFKL
type~\citep{BFKL1,BFKL2,BFKL3,BFKL4}:
\begin{eqnarray}
\int_0^{\mu^2} d{\bf q}_T^2 \ \Phi_q(x, {\bf q}_T^2, \mu^2) = F_q(x, \mu^2).
\label{uPDF2}
\end{eqnarray}

The normalization condition~(\ref{uPDF2}) can be achieved by
multiplying the right--hand--side of Eq.~(\ref{uPDF1}) by the
function $T_q(x, {\bf q}_T^2, \mu^2)$ usually refereed as a {\em
Sudakov form factor}:
\begin{eqnarray}
\Phi_q(x, {\bf q}_T^2, \mu^2) & = &
\frac{d}{d {\bf q}_T^2}
\left[ T_q(x, {\bf q}_T^2, \mu^2) \ F_q(x, {\bf q}_T^2) \right].
\label{uPDF3}
\end{eqnarray}
Substituting definition~(\ref{uPDF3}) into the Eq.~(\ref{uPDF2}),
one can obtain the boundary conditions on Sudakov form factor:
$T_q(x, {\bf q}_T^2 = 0, \mu^2) = 0$ and
$T_q(x, {\bf q}_T^2 = \mu^2, \mu^2) = 1$.
In the mKMRW model, quark uPDF are calculated by the formula~\cite{NS:2020}:
\begin{eqnarray}
\Phi_q(x, {\bf q}_T^2, \mu^2) & = &
\frac{\alpha_S({\bf q}_T^2)}{2 \pi} \frac{T_q(x, {\bf q}_T^2, \mu^2)}{{\bf q}_T^2}
\sum\limits_{j = g, q}
\int_x^1 dz \
P_{i j}(z) \ F_j\left( \frac{x}{z}, {\bf q}_T^2 \right)
\theta\left( \Delta({\bf q}_T^2, \mu^2) - z \right). \label{uPDF4}
\end{eqnarray}
The main difference from the original KMRW model~\cite{KMR,MRW} is
that in the mKMRW model~\cite{NS:2020} Sudakov form factor depends
also on $x$ variable, which is not true in the KMRW model. The exact
form of the Sudakov form factor defined in Eq.~(\ref{uPDF3}) has
been obtained in Ref.~\cite{NS:2020}:
\begin{eqnarray}
T_q(x, {\bf q}_T^2, \mu^2) & = &
\exp\left[ - \int_{{\bf q}_T^2}^{\mu^2}
\frac{dt'}{t'} \frac{\alpha_S(t')}{2 \pi}
\left( \tau_q(t', \mu^2) + \Delta \tau_q(x, t', \mu^2) \right)
\right]. \label{uPDF5}
\end{eqnarray}
The first term in the exponent~(\ref{uPDF5}) corresponds to the Sudakov
form factor in the KMRW model~\cite{KMR,MRW}, while the second $x$ dependent
term is added in the modified KMRW model~\cite{NS:2020} and ensure the
exact normalization on the collinear PDFs~(\ref{uPDF2}) for any $x$ and
${\bf q}_T^2$. Similarly, formula for the gluon uPDF can be obtained,
see Eq.~(22) in Ref.~\cite{NS:2020}.

Finally, replacing tree--level uPDFs in Eq.~(\ref{PRA0}) by the~(\ref{uPDF4}),
we arrive to the PRA factorization formula:
\begin{equation}
\sigma^{\rm (LO, \ PRA)}(p p \to \gamma X) \simeq
\Phi_q(x_+, {\bf q}_{T_1}^2, \mu^2) \otimes
{\cal H}^{\rm (LO, \ mMRK)}_\gamma(x_\pm, {\bf q}_{T_{1,2}}, \mu^2) \otimes
\Phi_{\bar q}(x_-, {\bf q}_{T_2}^2, \mu^2). \label{PRA}
\end{equation}
The factorization formula~(\ref{PRA}) may be proved for any set of
the initial Reggeized partons. Moreover, as it was
shown~\cite{HEF3}, NLO tree--level and loop corrections do not
destroy the form~(\ref{PRA}). So that the HEF factorization is valid
both in the LO approximation and NLO. The PRA beyond the LO was
considered in Refs.~\cite{NS:2017,NS:2018,MN:2020,MN:2021}.

\section{Direct photon production in the PRA}\label{sec:phot}

The tree--level partonic subprocesses for the direct photon production
of the orders ${\cal O}(\alpha^n \alpha_S^m)$ with $n + m \leq 2$ which
we consider in the work are the following:
\begin{eqnarray}
Q \ (q_1) + \bar Q \ (q_2) & \to & \gamma \ (q_3), \label{pr2} \\
Q \ (q_1) + R \ (q_2) & \to & \gamma \ (q_3) + q \ (q_4), \label{pr3} \\
Q \ (q_1) + \bar Q \ (q_2) & \to & \gamma \ (q_3) + g \ (q_4), \label{pr4}
\end{eqnarray}
where $R$ denote Reggeized gluon (Reggeon).
MEs for the subprocesses~(\ref{pr2})--(\ref{pr4})
were already obtained in Refs.~\cite{Saleev:2008,Saleev:2009,Saleev:2011}.

The LO subprocess~(\ref{pr2}) is not vanish since $q_{1, 2}^2 \neq 0$,
unlike to CPM case,
but doesn't correspond to the experimental setup, where cone distance
$r^2 = \Delta y^2 + \Delta \phi^2$ in the rapidity--azimuthal angle space
is finite, while in the case of the subprocess~(\ref{pr2}) the cone distance
is infinite.
Below we will show that this problem of the LO PRA $2 \to 1$ contribution
can be solved. The NLO${}^\star$ contributions from the subprocesses~(\ref{pr3})\
and~(\ref{pr4}) should be taken into account. The first one is IR safety in the
limit $|{\bf q}_{T_4}| \to 0$ since this is a LO quark--gluon scattering.

\begin{figure}[ht]
\centering
\includegraphics[scale=0.7]{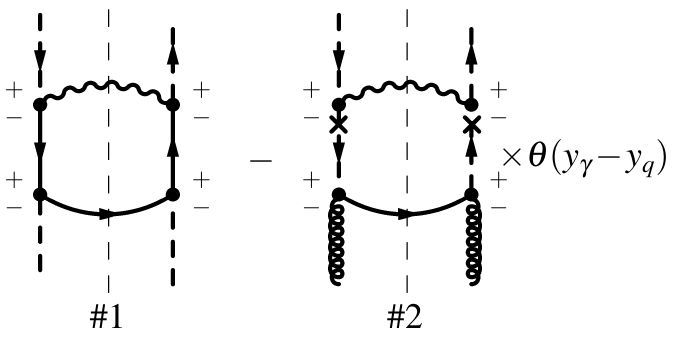} \\
\includegraphics[scale=0.7]{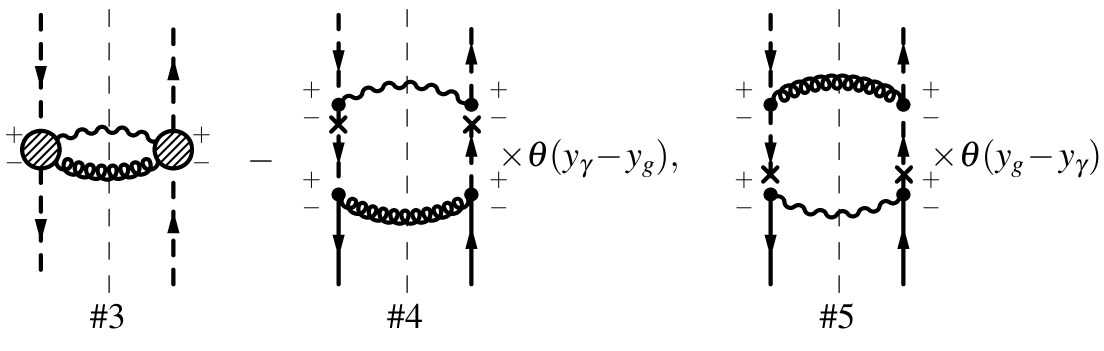}
\caption{mMRK subtraction scheme: subtraction diagram $\# 2$
corresponds to the subprocess~(\ref{pr3}) and $\# 4$ -- to the
subprocess~(\ref{pr4}).}\label{fig:3}
\end{figure}

Computations of the tree--level corrections in the HEF have
non--trivial problem of double counting between real corrections to
the hard--scattering coefficient and uPDFs: the NLO${}^\star$ cross
section $\sigma^{({\rm NLO}^\star)}$ is expressed through the
integral over the entire phase space of an additional parton $q$
in~(\ref{pr2}) and $g$ in~(\ref{pr3}), including the segment $|y_4|
< \infty$. In the PRA, there are only three rapidity regions:
forward $Y^+$ and backward $Y^-$ which consists partons from uPDFs
$\Phi(x_\pm, {\bf q}_{T_{1,2}}^2, \mu^2)$ with $y \to \pm \infty$ of
every parton, and the central region $Y$ of the production particles
with $|y| < \infty$, see Fig.~\ref{fig:1}. When additional parton
goes deeply to the $Y^\pm$ regions, it should be included into the
uPDF since in the PRA all the particles are produced in the one
rapidity cluster.

Cases with larger number of the rapidity clusters satisfies to other
MRK contributions. For example, contribution with Reggeized
propagator between photon and additional gluon is the zero
approximation of the BFKL Pomeron since Green function of the BFKL
equation has initial condition~\cite{BFKL4,Lipatov:10,Levin:12}: $G
\left( {\bf l}_T, {\bf l}_T', Y = 0 \right) = \delta^{(2)}\left(
{\bf l}_T - {\bf l}_T' \right)$, where ${\bf l}_T = {\bf q}_{T_1} -
{\bf q}_{T_3}$ and ${\bf l}_T' = {\bf q}_{T_4} - {\bf q}_{T_2}$, $Y
= |y(q_3) - y(q_4)|$ is a value of the rapidity gap. Such
contribution does not take into account the double counting between
hard--scattering coefficient and uPDF and should be considered as an
independent contribution corresponding to the case when particles
are produced in the different rapidity clusters, like
Mueller--Navelet dijets~\cite{Mueller:1986}.

The double counting subtraction term must satisfy three main requirements:
\begin{enumerate}
\item Subtraction term must be gauge--invariant;
\item The number of the rapidity regions must leave three;
\item The photon and additional parton must be strongly divided by the rapidities.
\end{enumerate}
Such situation may be achieved if we consider the propagator in the
$\hat t$--channel diagrams $\# 1$ and $\# 3$ in Fig.~\ref{fig:3},
which are parts of the full set of diagrams for
subprocesses~(\ref{pr3}) and~(\ref{pr4}) respectively, between
photon and additional parton to be Reggeized and opposite to the
photon (not connected by a vertex) parton on mass--shell. The $\hat
s$-- and $\hat u$--channel diagrams don't contribute to the
subtraction term since they are vanish in the Regge limit $\hat s
\sim - \hat u \gg - \hat t$. In Fig.~\ref{fig:3}, the set $\# 3$
also includes the $\hat u$--channel diagram, and the corresponding
squared subtraction amplitude is diagram $\# 5$, the last one not
should be added to the subtraction term for the
subprocess~(\ref{pr4}) since situation when $y_g > y_\gamma$ are
taken into account into the permutation $(+ \leftrightarrow -)$,
which also implements permutation $(\hat t \leftrightarrow \hat u)$.

The subtraction amplitudes for the subprocesses~(\ref{pr3}) and~(\ref{pr4})
in the spinor representation, see diagrams $\# 2$ and $\# 4$
in Fig.~\ref{fig:3}, reads:
\begin{eqnarray}
{\cal A}^{\mu\nu}_a (Q g \to \gamma q) & = &
i e g \, T_a \,
\bar u(q_4) \,
\gamma^\mu_{(+)}(q, -q_4) \, D^+_{\times}(q) \,
\Gamma^\nu_{(+-)}(q_1, -q_3) \, u(q_{L_1}), \label{am1} \\
{\cal A}^{\mu\nu}_a (Q \bar q \to \gamma g) & = &
i e g \, T_a \,
\bar u(q_2) \,
\gamma^\mu_{(+)}(q, -q_4) \, {}_{\times} D^+(q) \,
\Gamma^\nu_{(+-)}(q_1, -q_3) \, u(q_{L_1}), \label{am2}
\end{eqnarray}
where $q = q_1 - q_3$ and $q_2 \simeq \left( q_2^- / 2 \right) n_+$,
rapidity ordering $\theta$--functions $\theta\left( y(q_3) - y(q_4)
\right)$ are omitted and will be restored later. It is easy to check
that in such a way defined subtraction amplitudes~(\ref{am1})
and~(\ref{am2}) for the subprocesses~(\ref{pr3}) and~(\ref{pr4})
respectively satisfies Slavnov--Teylor identities for the on--shell
gluon and photon due to the gauge--invariance of the effective
vertices~\cite{Lipatov:1,Lipatov:2}: $q_{2 \mu, 3 \nu} {\cal A}^{\mu
\nu}_a(Q g \to \gamma q) = 0$ and $q_{3 \nu, 4 \mu} {\cal A}^{\mu
\nu}_a(Q \bar q \to \gamma g) = 0$.

First it was first shown in Ref.~\cite{NS:2016} that squared amplitudes~(\ref{am1})
and~(\ref{am2}) may be factorized into the IR ${\bf q}_{T_1}^2 \to 0$ safety
factor, which is vanish in the MRK limit $(- \hat t) / \hat s \to 0$,
DGLAP splitting function, and the $2 \to 1$ ME for the subprocess
$Q ({\bf q}_{T_1}) \, \bar Q ({\bf q}_{T}) \to \gamma$ defined in Eq.~(\ref{2To1}):
\begin{eqnarray}
\overline{\mid {\cal A}^{\rm (mMRK, \ sub.)}
(Q g \to \gamma q) \mid^2} & = &
g^2 \, \frac{2 \, C_F \, N_c}{C_A \, (N_c^2 - 1)}
\frac{\hat s + {\bf q}_{T_1}^2}{\hat s + \hat t + {\bf q}_{T_1}^2}
\frac{T_F^{-1} P_{q g}(z)}{z \, (- \hat t)} \,
\theta\left( y(q_3) - y(q_4) \right)
\nonumber \\ & \times &
\overline{\mid {\cal A}^{\rm (MRK)}
(Q ({\bf q}_{T_1}) \, \bar Q ({\bf q}_{T}) \to \gamma) \mid^2}, \label{Sub1} \\
\overline{\mid {\cal A}^{\rm (mMRK, \ sub.)}
(Q \bar q \to \gamma g) \mid^2} & = &
g^2 \, \frac{2 \, C_F}{C_A}
\frac{\hat s + {\bf q}_{T_1}^2}{\hat s + \hat t + {\bf q}_{T_1}^2}
\frac{C_F^{-1} P_{q q}(z)}{z \, (- \hat t)} \,
\theta\left( y(q_3) - y(q_4) \right)
\nonumber \\ & \times &
\overline{\mid {\cal A}^{\rm (MRK)}
(Q ({\bf q}_{T_1}) \, \bar Q ({\bf q}_{T}) \to \gamma) \mid^2}, \label{Sub2}
\end{eqnarray}
here $\hat s = q_1^+ q_2^- / 2$ and $\hat t = - q_3^- q_4^+ - {\bf q}_T^2$ are the
Mandelstam variables, momenta fraction $z = q_4^- / q_2^-$, and
$P_{q g}(z) = T_F \left[ z^2 + (1 - z)^2 \right]$ with $T_F = 1 / 2$.

To derive the subtraction term cross section formula, first one may take
the limit ${\bf q}_{T_2}^2 \to 0$ in the hard scattering coefficient. Such
procedure is IR safe since in the PRA Reggeons $R_\pm$ have initial state
factors $q^\pm / (2 |{\bf q}_{T_{1,2}}|)$ regularize IR divergences
$|{\bf q}_{T_{1,2}}| \to 0$, see Eq.~(4.15) in Ref.~\cite{HEF2}.
Then one may integrate uPDF $\Phi(x_-, {\bf q}_{T_2}^2, \mu^2)$ over the
${\bf q}_{T_2}$ with the help of exact normalization condition, see
Eq.~(\ref{uPDF2}). The subtraction term cross section then reads:
\begin{eqnarray}
\sigma^{\rm (mMRK, \ sub.)} \simeq
\Phi(x_+, {\bf q}_{T_1}^2, \mu^2) \otimes
{\cal H}^{\rm (mMRK, \ sub.)}(x_\pm, {\bf q}_{T_{1}}, \mu^2)
\otimes F(x_-, \mu^2). \label{SubT}
\end{eqnarray}
Finally, we can match the LO contribution with the NLO${}^\star$ tree--level
corrections and the relevant double counting subtraction as follows:
\begin{eqnarray}
\sigma =
\sigma^{\rm (LO, \ PRA)} + \sigma^{({\rm NLO}^\star, \ {\rm PRA})} -
\sigma^{\rm (mMRK, \ sub.)}. \label{CStot}
\end{eqnarray}

\begin{figure}[ht]
\centering
\includegraphics[scale=0.25]{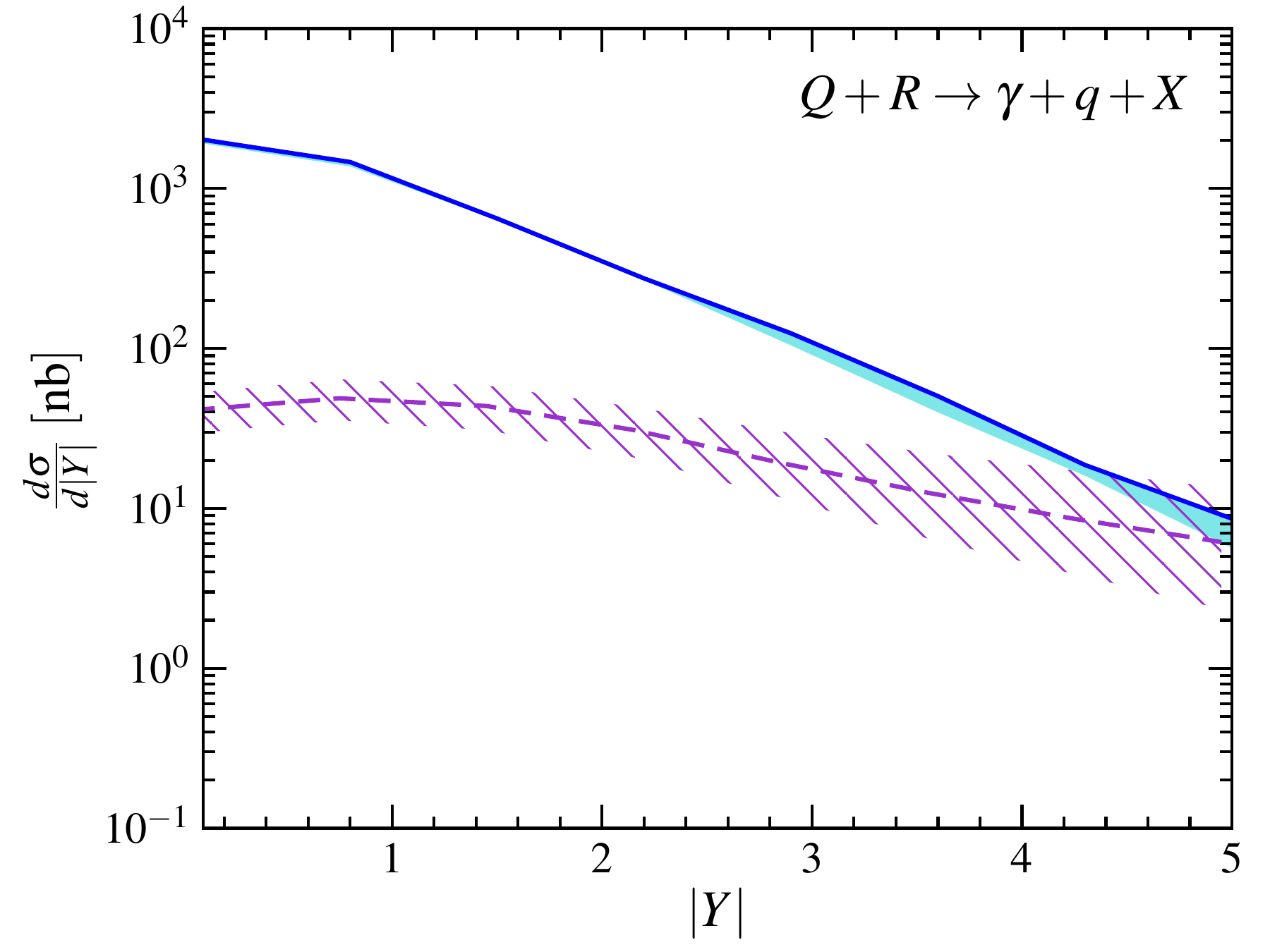}
\includegraphics[scale=0.25]{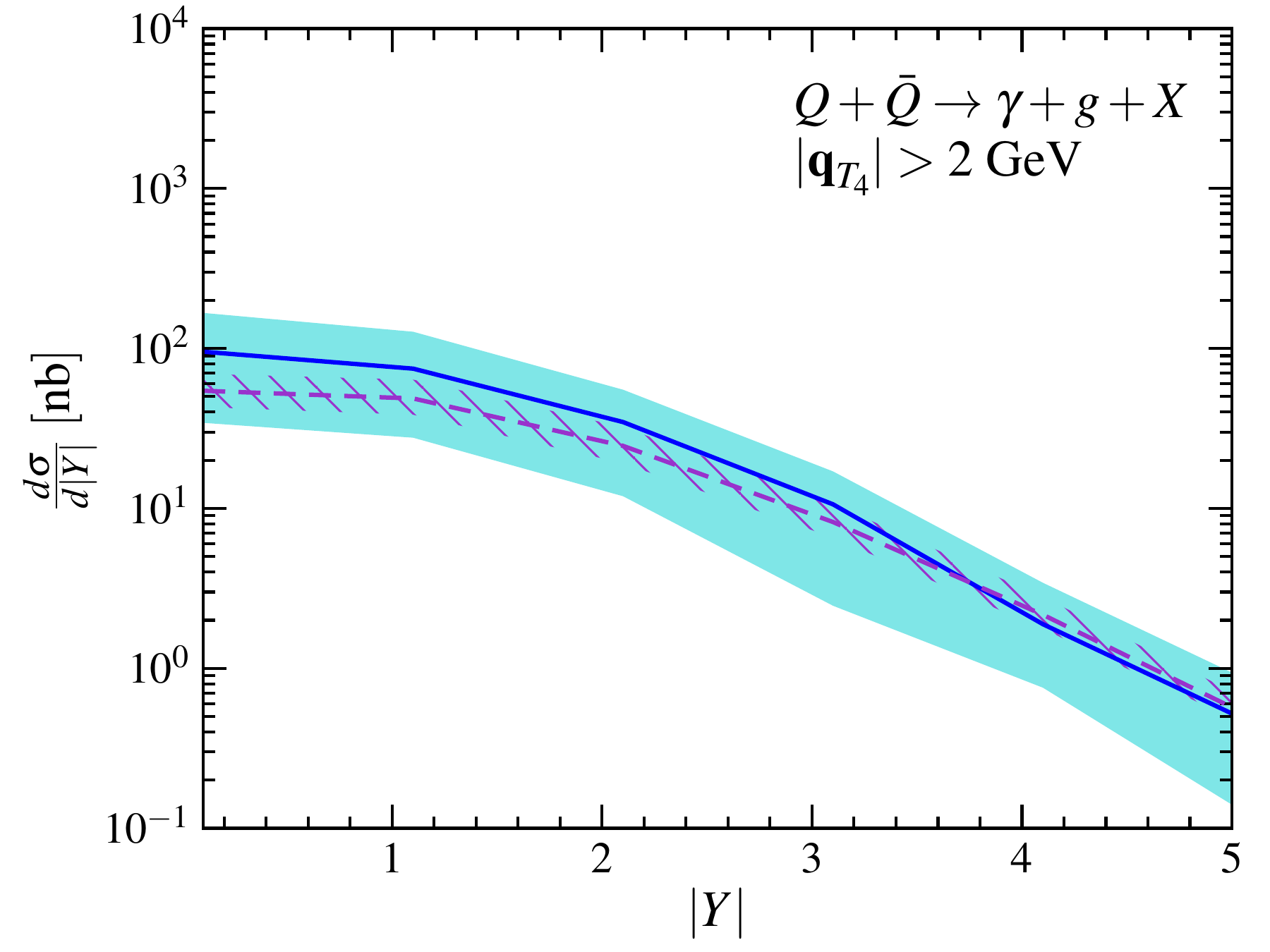}
\caption{Rapidity difference spectra of the two subprocesses: left
panel is subprocess~(\ref{pr3}), right panel is the
subprocess~(\ref{pr4}). Blue solid line is the NLO${}^\star$
contribution, violet dashed line is the corresponding double
counting subtraction.}\label{fig:4}
\end{figure}

To show self--consistency of the mMRK subtraction scheme, in
Fig.~\ref{fig:3} we plot spectra on rapidity difference $Y =
y^\gamma - y^{q, g}$ of the two subprocesses~(\ref{pr3})
and~(\ref{pr4}) since they are the most specific for the subtraction
scheme. In the case of the subprocess~(\ref{pr4}) we numerically
regularize IR divergence using cutoff $|{\bf q}_{T_4}| > 2$ GeV. In
both cases, the NLO${}^\star$ contributions coincides with the
relevant mMRK subtraction as it should be. While in the
subprocess~(\ref{pr3}) subtraction becomes large at $|Y| \geq 4$, in
the subprocess~(\ref{pr4}) the subtraction is large even at
intermediate values of $|Y|$. The last one shows the
self--consistency of the PRA first demonstrated in
Ref.~\cite{NS:2016}.

At the stage of the numerical calculations we used the {\tt Suave}
Monte--Carlo algorithm implemented in the {\tt CUBA} library~\citep{Cuba}.
The results of calculations of the subprocesses~(\ref{pr2})--(\ref{pr4}) are
also compared with the results obtained using MC parton level event generator
{\tt KaTiE}~\cite{katie}.

\section{Results}\label{sec:res}

In the experiment, it is difficult task to extract the contribution
of the direct photons, so the contribution of prompt or isolated
photons is usually studied. The contribution of the {\em prompt}
photons includes direct photons contribution and fragmentation
production. To estimate the fragmentation contribution, one may
introduce relevant fragmentation functions (FFs) of parton to
photon, but this method have some problems with double counting
between FF and NLO${}^\star$ corrections. To suppress the
contribution of the fragmentation photons and obtain a good
approximation for the direct photons contribution, the {\em
isolation cone condition} is usually introduced:
\begin{eqnarray}
r & = & \sqrt{\Delta y^2 + \Delta \phi^2} \geq R^{\rm (iso)}. \label{cone}
\end{eqnarray}
This condition requires no hadrons with $E_T > E_T^{\rm (iso)}$ inside the
photon cone $r \leq R^{\rm (iso)}$ in the rapidity--azimuthal angle space,
parameters $R^{\rm (iso)}$ and $E_T^{\rm (iso)}$ are taken from the experimental setup.
To estimate the fragmentation photons contribution, one may use
the Frixione modification of the cone condition:
\begin{eqnarray}
E_T & \leq & E_T^{\rm (iso)} \ \chi(r ; n), \qquad
\chi(r ; n) = \left( \frac{1 - \cos r}{1 - \cos R^{\rm (iso)}} \right)^n
\label{Frixione}
\end{eqnarray}
with $n \geq 1/2$. In the measurements of ATLAS Collaboration, a
$p_T^\gamma$--dependent isolation is used:
$E_T^{\rm (iso)} < 4.2 \cdot 10^{-3} \times p_T^\gamma + 4.8 \ {\rm GeV}$.
We have found that impact of the Frixione cone condition~(\ref{Frixione}) is
negligibly small and standard isolation cone condition~(\ref{cone}) is a good
approximation of the direct contribution we are interested in.

\renewcommand{\arraystretch}{1.}
\renewcommand{\tabcolsep}{.2cm}

\begin{table}[t]
\centering

\begin{tabular}{c c c c}

\hline\hline

\multicolumn{4}{c} {D$\emptyset$~\cite{D0:2005} $\colon$ $\sqrt{S} =
1.96$ TeV, $R^{\rm (iso)} = 0.4$,
$23 < p_T^\gamma < 300$ GeV, $|y^\gamma| < 0.9$} \\

\hline

\multicolumn{4}{c} {ATLAS~\cite{ATLAS:2016} $\colon$ $\sqrt{S} = 8$
TeV, $R^{\rm (iso)} = 0.4$,
$25 < p_T^\gamma < 1500$ GeV} \\

$Y_1 \colon |y^\gamma| < 0.6$ & $Y_2 \colon 0.6 < |y^\gamma| < 1.37$
&
$Y_3 \colon 1.56 < |y^\gamma| < 1.81$ & $Y_2 \colon 1.81 < |y^\gamma| < 2.37$ \\

\hline

\multicolumn{4}{c} {ATLAS~\cite{ATLAS:2019} $\colon$ $\sqrt{S} = 13$
TeV, $R^{\rm (iso)} = 0.4$,
$125 < p_T^\gamma < 2500$ GeV} \\

$Y_1 \colon |y^\gamma| < 0.6$ & $Y_2 \colon 0.6 < |y^\gamma| < 1.37$
&
$Y_3 \colon 1.56 < |y^\gamma| < 1.81$ & $Y_2 \colon 1.81 < |y^\gamma| < 2.37$ \\

\hline

\multicolumn{4}{c} {ATLAS~\cite{ATLAS:2023} $\colon$ $\sqrt{S} = 13$
TeV, $R^{\rm (iso)} = 0.2$ and $R^{\rm (iso)} = 0.4$,
$250 < p_T^\gamma < 2500$ GeV} \\

$Y_1 \colon |y^\gamma| < 0.6$ & $Y_2 \colon 0.6 < |y^\gamma| < 0.8$
&
$Y_3 \colon 0.8 < |y^\gamma| < 1.37$ & $Y_4 \colon 1.56 < |y^\gamma| < 1.81$ \\
& $Y_5 \colon 1.81 < |y^\gamma| < 2.01$ & $Y_6 \colon 2.01 < |y^\gamma| < 2.37$ \\

\hline

\multicolumn{4}{c} {CMS~\cite{CMS:2018} $\colon$ $\sqrt{S} = 13$
TeV, $R^{\rm (iso)} = 0.3$,
$190 < p_T^\gamma < 1000$ GeV} \\

$Y_1 \colon |y^\gamma| < 0.8$ & $Y_2 \colon 0.8 < |y^\gamma| < 1.44$
&
$Y_3 \colon 1.57 < |y^\gamma| < 2.1$ & $Y_2 \colon 2.1 < |y^\gamma| < 2.5$ \\

\hline

\multicolumn{4}{c} {PHENIX~\cite{PHENIX:2012} $\colon$ $\sqrt{S} =
200$ GeV, $R^{\rm (iso)} = 0.5$,
$5 < p_T^\gamma < 26$ GeV, $|y^\gamma| < 0.35$} \\

\hline

\multicolumn{4}{c} {UA6~\cite{UA6:1998} $\colon$ $\sqrt{S} = 24.3$
GeV,
$4.1 < p_T^\gamma < 7.7$ GeV, $0.1 < y^\gamma < 0.9$} \\

\hline\hline

\end{tabular}

\caption{\centering Phase space cuts of the isolated photon
production measurements in the different
experiments~\cite{D0:2005,ATLAS:2016,ATLAS:2019,ATLAS:2023,CMS:2018,
PHENIX:2012,UA6:1998}.}\label{tab:1}

\end{table}

\begin{figure}[ht]
\centering
\includegraphics[scale=0.3]{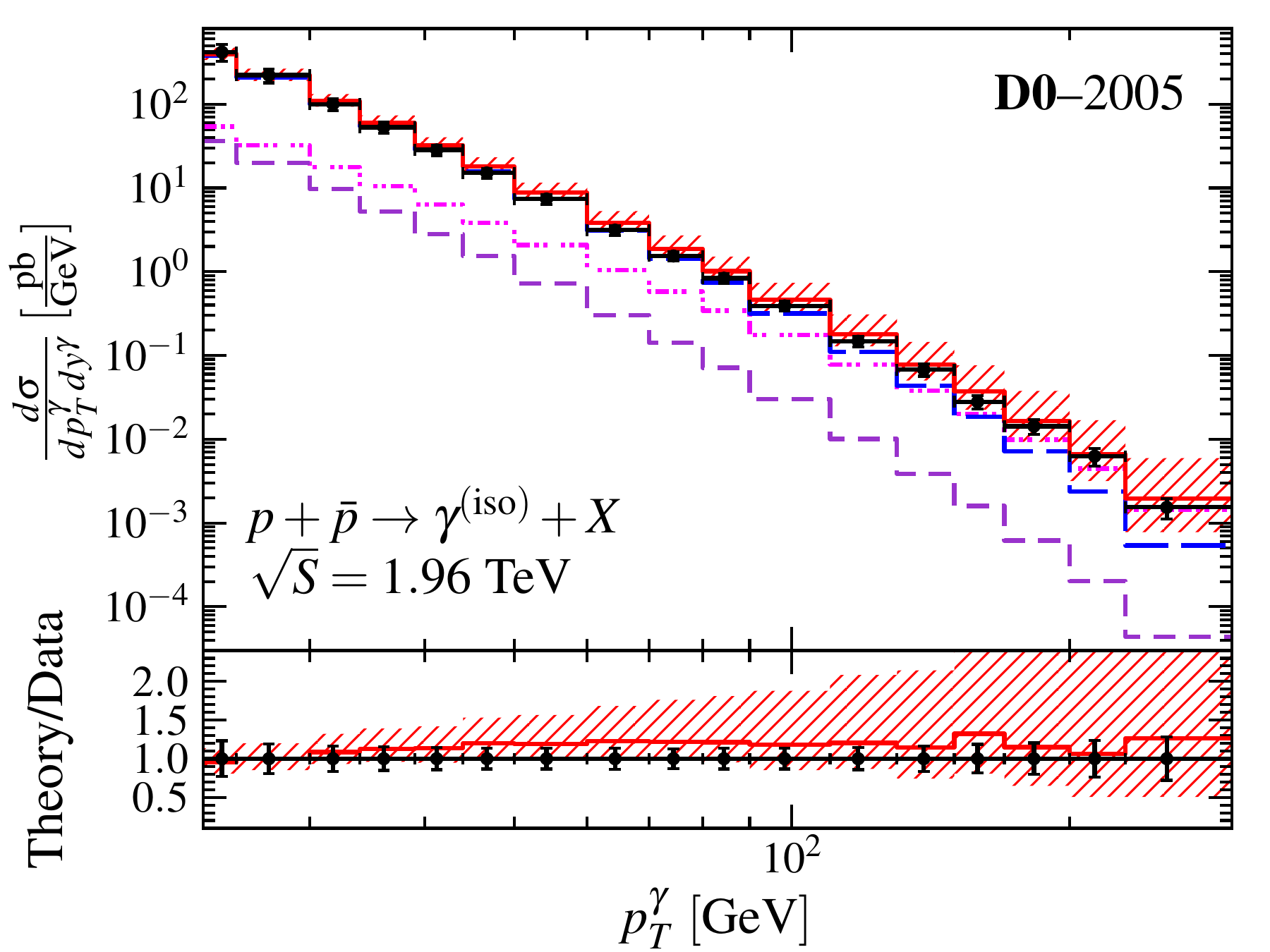}
\caption{ Transverse energy spectra of the isolated photon
production at the $\sqrt{S} = 1.96$ TeV. Solid red line with hatch
is the total contribution after subtraction~(\ref{CStot}), dashed
blue line and dash--dotted magenta lines are the
NLO${}^\star$~(\ref{pr3}) and LO~(\ref{pr2}) contributions
respectively, and loosely dashed violet line is the subtraction
term~(\ref{SubT}). The data are from D$\emptyset$
Collaboration~\cite{D0:2005}.}\label{fig:5}
\end{figure}

\begin{figure}
\centering
\includegraphics[scale=0.2]{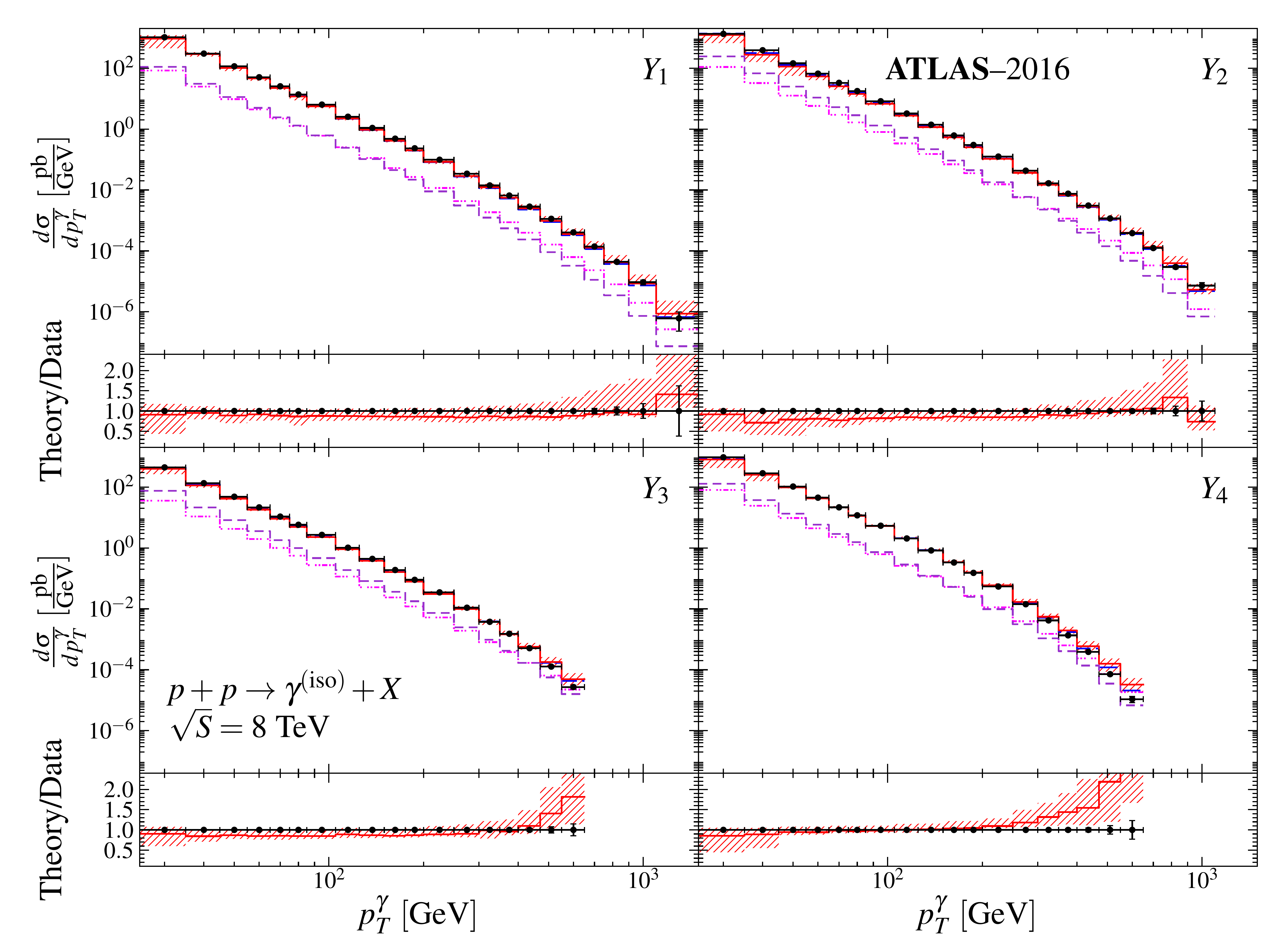}
\caption{ Transverse energy spectra of the isolated photon
production at the $\sqrt{S} = 8$ TeV. The notations of the curves
are the same as in Fig.~\ref{fig:5}. The data are from ATLAS
Collaboration~\cite{ATLAS:2016}.}\label{fig:6}
\end{figure}

\begin{figure}
\centering
\includegraphics[scale=0.2]{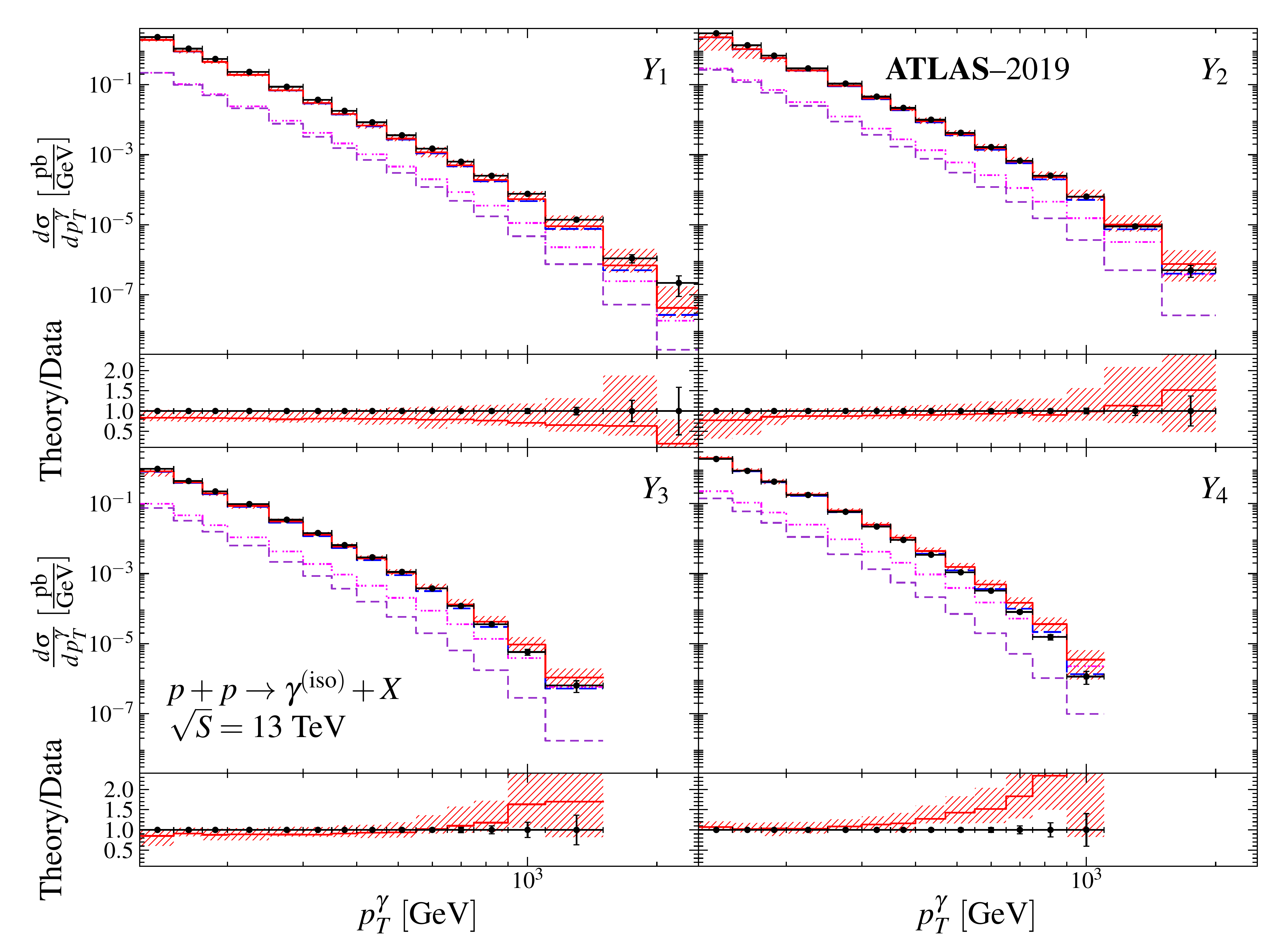}
\caption{ Transverse energy spectra of the isolated photon
production at the $\sqrt{S} = 13$ TeV. The notations of the curves
are the same as in Fig.~\ref{fig:5}. The data are from ATLAS
Collaboration~\cite{ATLAS:2019}.}\label{fig:7}
\end{figure}

\begin{figure}
\centering
\includegraphics[scale=0.2]{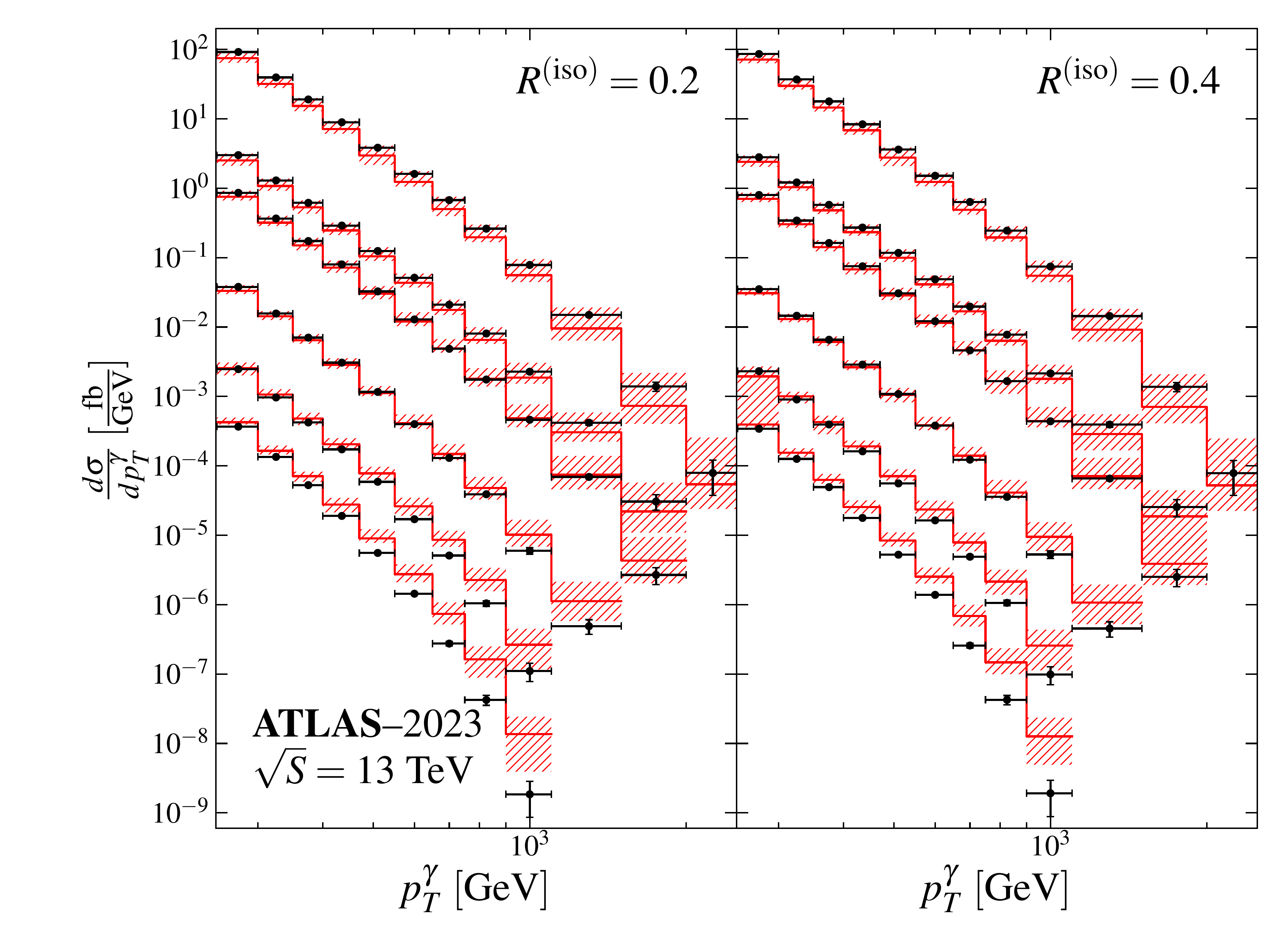}
\caption{ Transverse energy spectra of the isolated photon
production at the $\sqrt{S} = 13$ TeV. The notations of the curves
are the same as in Fig.~\ref{fig:5}. The data are from ATLAS
Collaboration~\cite{ATLAS:2023}.}\label{fig:8}
\end{figure}

\begin{figure}
\centering
\includegraphics[scale=0.2]{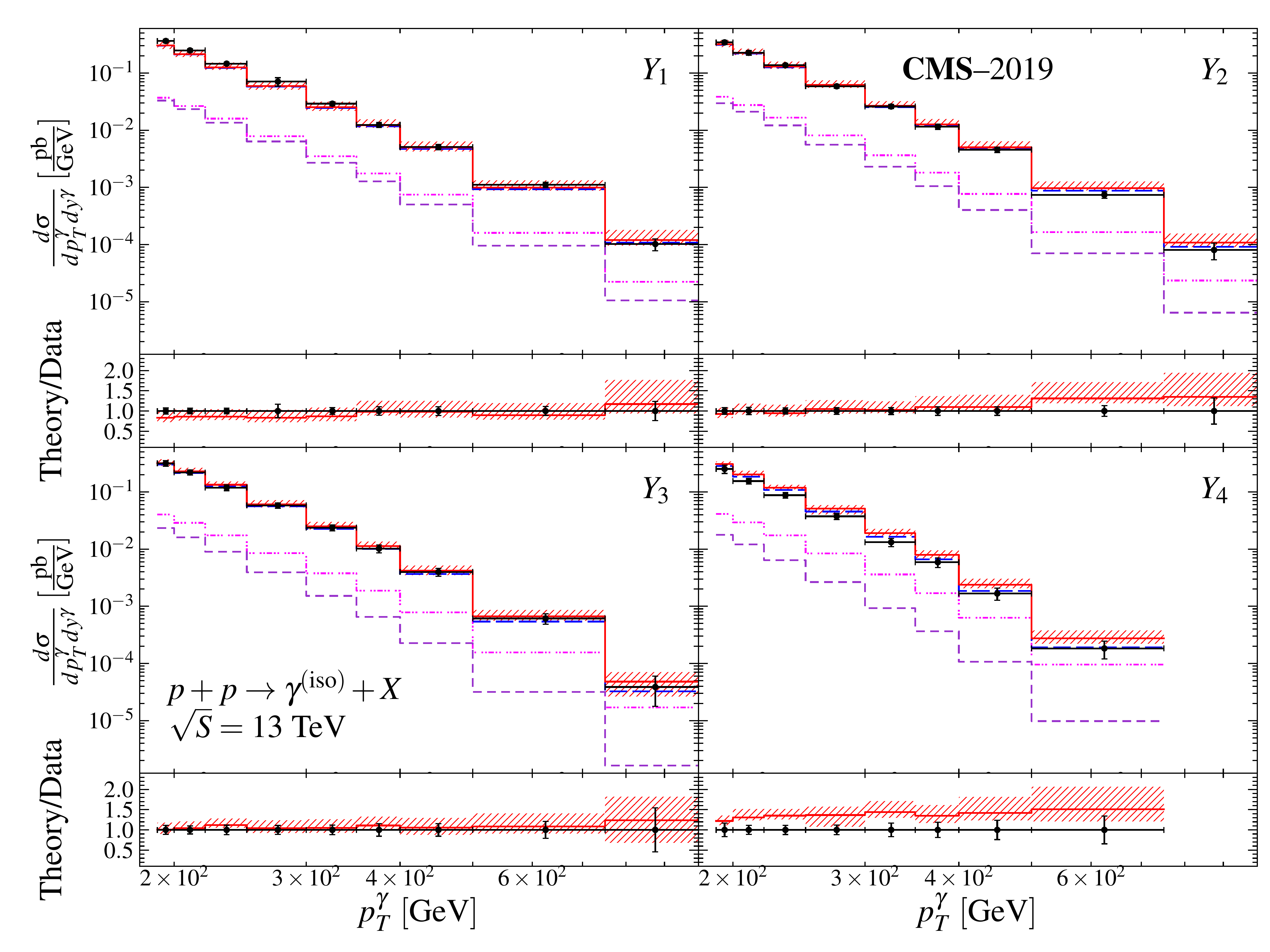}
\caption{ Transverse energy spectra of the isolated photon
production at the $\sqrt{S} = 13$ TeV. The notations of the curves
are the same as in Fig.~\ref{fig:5}. The data are from CMS
Collaboration~\cite{CMS:2018}.}\label{fig:9}
\end{figure}

In Table~\ref{tab:1} we collected phase space cuts of measurements from
different experiments and wide energy range
$\sqrt{S} = 24.3 \ {\rm GeV} - 13 \ {\rm TeV}$, with which we compared our
predictions, see Figs.~\ref{fig:5}--\ref{fig:9}.
The notations of the curves are explained
in the caption to Fig.~\ref{fig:5}. To estimate high order corrections,
we choose the central value of the hard scale $\mu$ to be equal
$p_T^\gamma / 2$, the
variation by a factors $\xi = 2^{\pm 1}$ from the central value gives
upper and lower estimates of the cross--section, i.e.
$\sigma(\mu) \in \left[ \sigma(2 \mu), \sigma(\mu / 2) \right]$.

First we discuss the isolated photon production at high energies
$\sqrt{S} = 2 - 13$ TeV, see Figs.~\ref{fig:5}--\ref{fig:8}. The
LO~(\ref{pr2}) and NLO${}^\star$~(\ref{pr3}) contributions are shown
separately, as well as mMRK subtraction term~(\ref{SubT}) with
ME~(\ref{Sub1}). The subtraction term~(\ref{SubT}) mostly coincides
with the LO contribution~(\ref{pr2}) due to the
relation~(\ref{Sub1}), exclude region of high $p_T^\gamma$, and only
the NLO${}^\star$ contribution~(\ref{pr3}) remains. Our predictions
agrees with data well up to $p_T^\gamma / \sqrt{S} \simeq 0.2$. The
only disagreement is found at high $p_T^\gamma$ in the rapidity
regions $Y_{i \geq 3}$, see Table~\ref{tab:1} for explanation, where
our predictions overestimate data by about a factor of $2$, but also
the MRK condition $\mu \ll \sqrt{S}$ under the question in this
regions. In Ref.~\cite{ATLAS:2023} of ATLAS Collaboration, two sets
of transverse momenta spectra at two different values of the
isolation cone radius have been extracted, see Table~\ref{tab:1}. We
do not find any special discussion about it, since our predictions
agrees with data approximately equally in both cases, see
Fig~\ref{fig:8}.

\begin{figure}
\centering
\includegraphics[scale=0.3]{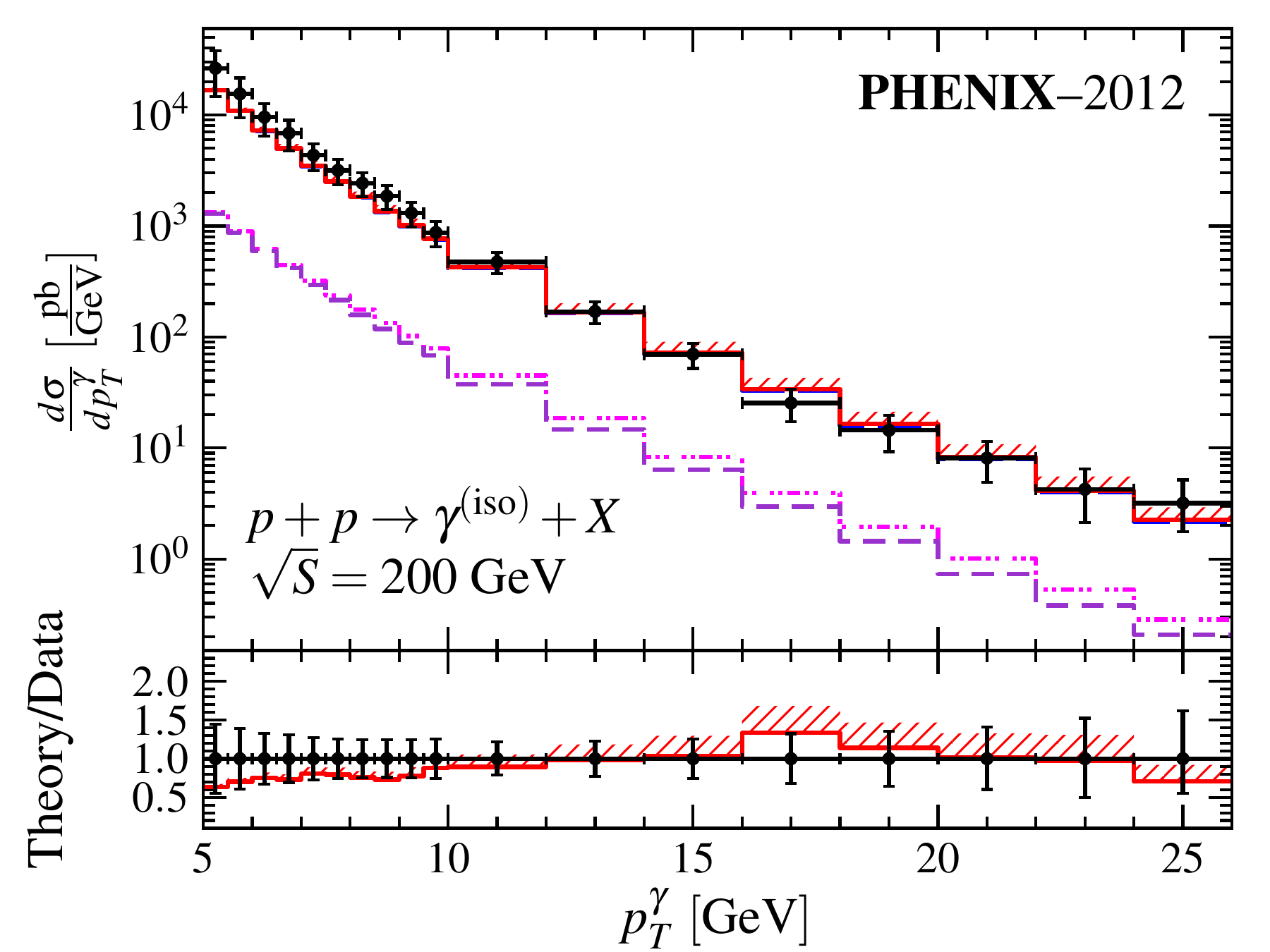}
\caption{ Transverse energy spectra of the isolated photon
production at the $\sqrt{S} = 200$ GeV. The notations of the curves
are the same as in Fig.~\ref{fig:5}. The data are from PHENIX
Collaboration~\cite{PHENIX:2012}.}\label{fig:10}
\end{figure}

\begin{figure}
\centering
\includegraphics[scale=0.3]{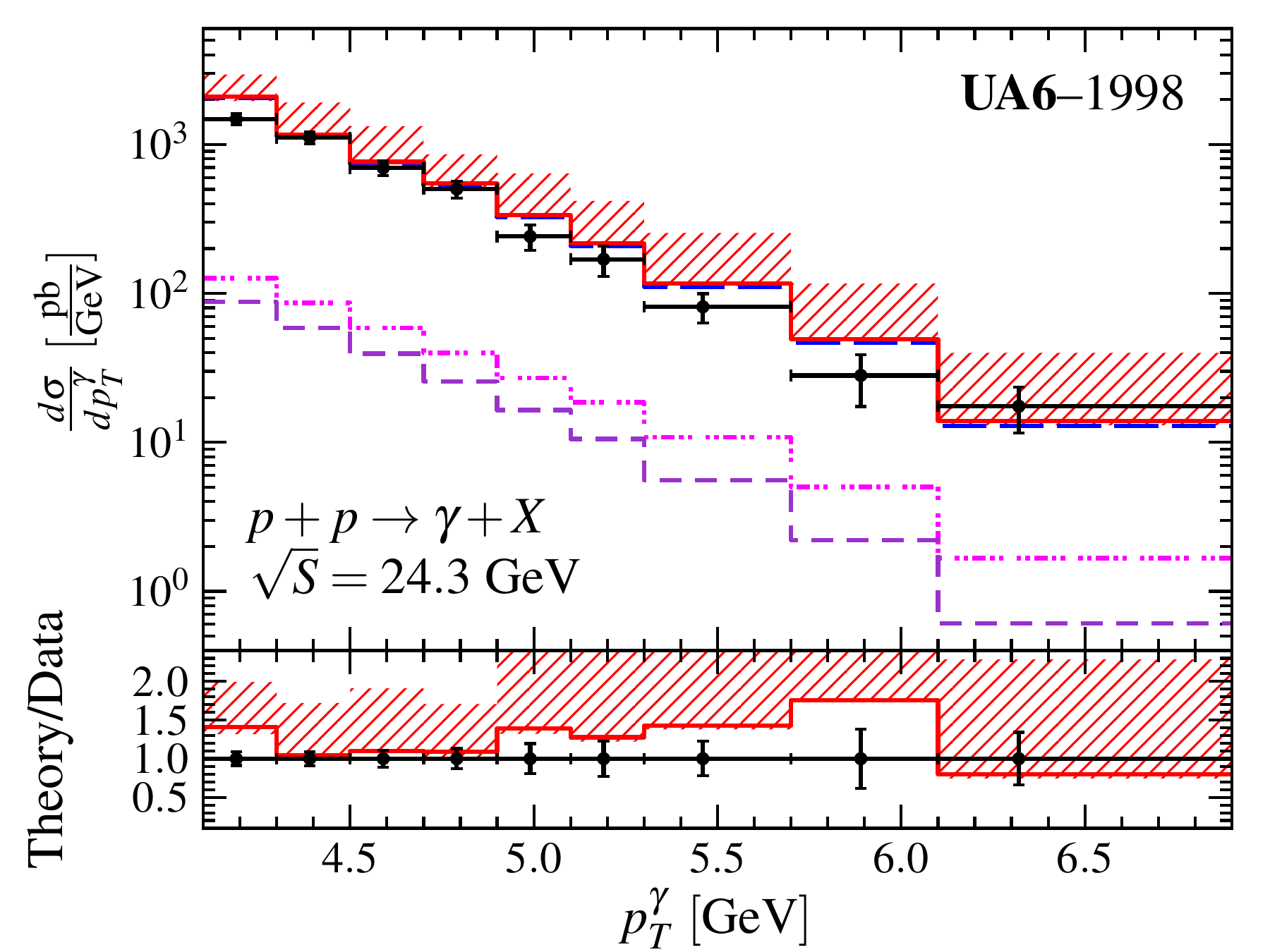}
\caption{ Transverse energy spectra of the isolated photon
production at the $\sqrt{S} = 24.3$ GeV. The notations of the curves
are the same as in Fig.~\ref{fig:5}. The data are from UA6
Collaboration~\cite{UA6:1998}.}\label{fig:11}
\end{figure}

\begin{figure}
\centering
\includegraphics[scale=0.3]{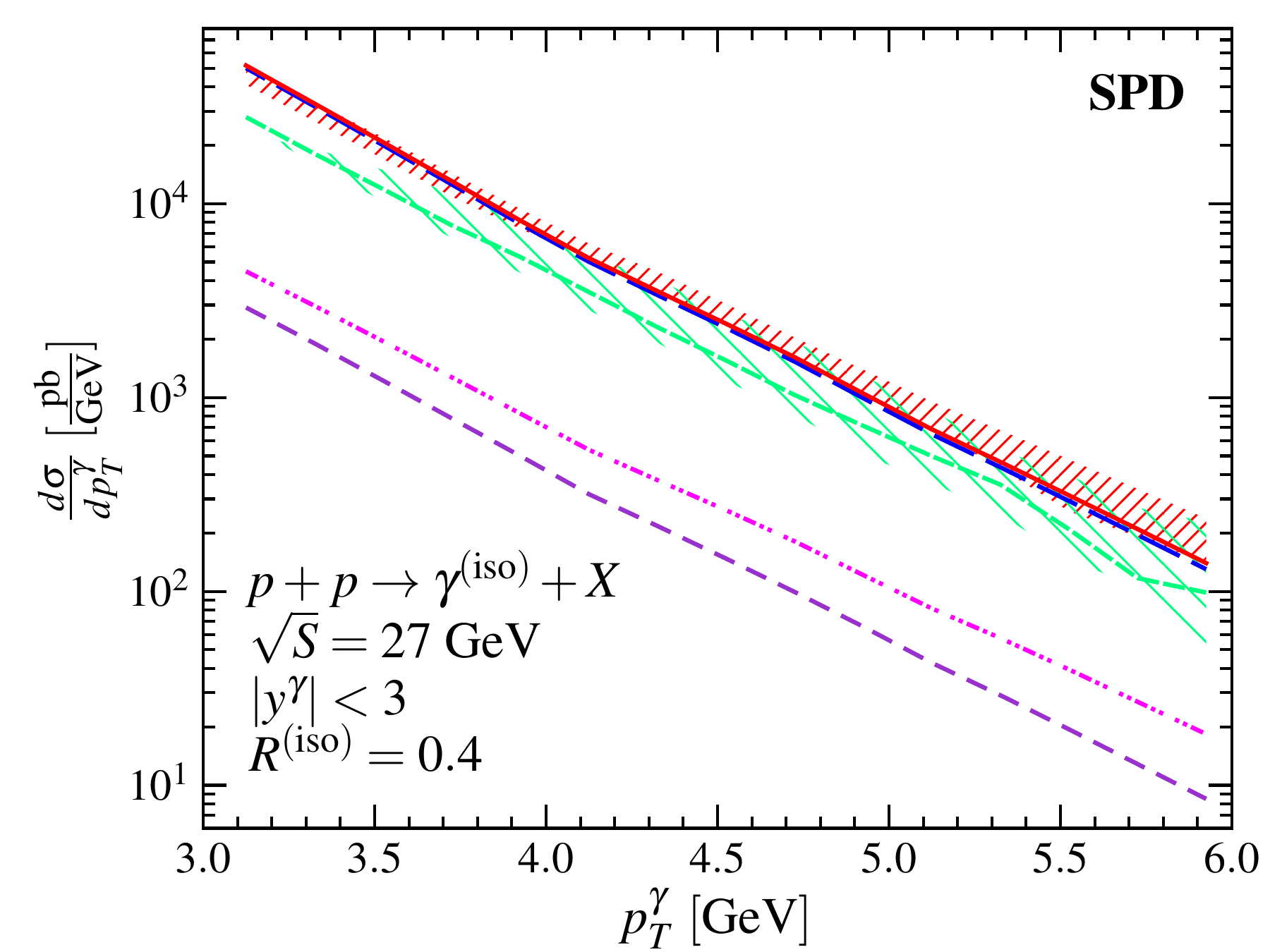}
\caption{ Transverse energy spectra of the isolated photon
production at the $\sqrt{S} = 27$ GeV. The notations of the curves
are the same as in Fig.~\ref{fig:5}, dashed green line with hatch is
the LO CPM predictions.}\label{fig:12}
\end{figure}

Then we tried to extrapolate our predictions to the low energy
region, see Figs.~\ref{fig:10}--\ref{fig:11}. At the energy
$\sqrt{S} = 200$ GeV of PHENIX~\cite{PHENIX:2012} experiment our
predictions are still in good agreement with data and the
relationship between the subtraction term~(\ref{SubT}) and the LO
contribution~(\ref{pr2}) is preserved. Our predictions a little bit
underestimate data at low $p_T^\gamma < 10$ GeV, but do not
intersect experimental errors. At the energy $\sqrt{S} = 24.3$ GeV
of UA6 experiment~\cite{UA6:1998} our predictions at the central
value of the hard scale $\mu = p_T^\gamma / 2$ approximately agree
with data. Hard scale variation gives only the increasing cross
section in this case, what is not found in all other cases. Taking
into account good agreement between our calculations using NLO
${}^\star$ approximation of the PRA and the data at high and low
energies, in Fig.~\ref{fig:12} we present our predictions for the
isolated photon transverse momenta spectra in the future SPD
experiment~\cite{SPD} at $\sqrt{S} = 27$ GeV with $|y^\gamma| < 3$.
We found the dependence on the $R^{\rm (iso)}$ parameter defined in
Eq.~(\ref{cone}) to be small in the NLO${}^\star$ PRA calculations
at the SPD NICA kinematical range, so we present predictions only
for $R^{\rm (iso)} = 0.4$. We also compare the results of our
NLO${}^\star$ PRA calculations with the LO CPM results, the last one
is shown in Fig.~\ref{fig:12} as dashed green line with hatch. Our
NLO${}^\star$ PRA predictions are higher than LO CPM predictions, a
similar situation has been observed many times before in the
different processes, see, for
example,~\cite{NS:2014,NS:2016,NS:2020}.

\section{Conclusions}\label{sec:conc}

In the paper, the production of the single isolated photon within the framework
of the NLO${}^\star$ approximation of the Parton Reggeization Approach is studied.
The new modified Multi--Regge Kinematics double counting subtraction scheme is
proposed.
We have obtained a quite satisfactory description of the transverse momentum
spectra at wide energy range up to $p_T^\gamma / \sqrt{S} \simeq 0.2 - 0.3$.
A special discussion of the direct photon production at low energies
$\sqrt{S} \simeq 24 - 200$ GeV was done, as well as the prediction for the
future SPD NICA experiment.
We demonstrated that NLO tree--level correction $Q \bar Q \to \gamma g$ to the
already existed LO subprocesses $Q \bar Q \to \gamma$ is small after subtraction
procedure, this shows the self--consistency of the Parton Reggeization Approach.

\section*{Acknowledgments}

We are grateful to Maxim Nefedov for a fruitful discussion on the
direct photon production in the High--Energy Factorization approach.
The work is supported by the Foundation for the Advancement of
Theoretical Physics and Mathematics BASIS, grant No. 24--1--1--16--5
and by the grant of the Ministry of Science and Higher Education of
the Russian Federation, No. FSSS--2024--0027.

\newpage

\bibliography{References}

\begin{thebibliography}{51}%
\makeatletter
\providecommand \@ifxundefined [1]{%
 \@ifx{#1\undefined}
}%
\providecommand \@ifnum [1]{%
 \ifnum #1\expandafter \@firstoftwo
 \else \expandafter \@secondoftwo
 \fi
}%
\providecommand \@ifx [1]{%
 \ifx #1\expandafter \@firstoftwo
 \else \expandafter \@secondoftwo
 \fi
}%
\providecommand \natexlab [1]{#1}%
\providecommand \enquote  [1]{``#1''}%
\providecommand \bibnamefont  [1]{#1}%
\providecommand \bibfnamefont [1]{#1}%
\providecommand \citenamefont [1]{#1}%
\providecommand \href@noop [0]{\@secondoftwo}%
\providecommand \href [0]{\begingroup \@sanitize@url \@href}%
\providecommand \@href[1]{\@@startlink{#1}\@@href}%
\providecommand \@@href[1]{\endgroup#1\@@endlink}%
\providecommand \@sanitize@url [0]{\catcode `\\12\catcode `\$12\catcode
  `\&12\catcode `\#12\catcode `\^12\catcode `\_12\catcode `\%12\relax}%
\providecommand \@@startlink[1]{}%
\providecommand \@@endlink[0]{}%
\providecommand \url  [0]{\begingroup\@sanitize@url \@url }%
\providecommand \@url [1]{\endgroup\@href {#1}{\urlprefix }}%
\providecommand \urlprefix  [0]{URL }%
\providecommand \Eprint [0]{\href }%
\providecommand \doibase [0]{https://doi.org/}%
\providecommand \selectlanguage [0]{\@gobble}%
\providecommand \bibinfo  [0]{\@secondoftwo}%
\providecommand \bibfield  [0]{\@secondoftwo}%
\providecommand \translation [1]{[#1]}%
\providecommand \BibitemOpen [0]{}%
\providecommand \bibitemStop [0]{}%
\providecommand \bibitemNoStop [0]{.\EOS\space}%
\providecommand \EOS [0]{\spacefactor3000\relax}%
\providecommand \BibitemShut  [1]{\csname bibitem#1\endcsname}%
\let\auto@bib@innerbib\@empty
\bibitem [{\citenamefont {Abazov}\ \emph {et~al.}(2006)\citenamefont {Abazov}
  \emph {et~al.}}]{D0:2005}%
  \BibitemOpen
  \bibfield  {author} {\bibinfo {author} {\bibfnamefont {V.~M.}\ \bibnamefont
  {Abazov}} \emph {et~al.} (\bibinfo {collaboration} {D0}),\ }\bibfield
  {title} {\bibinfo {title} {{Measurement of the isolated photon cross section
  in $p \bar{p}$ collisions at $\sqrt{s}$ = 1.96-TeV}},\ }\href
  {https://doi.org/10.1016/j.physletb.2006.04.048} {\bibfield  {journal}
  {\bibinfo  {journal} {Phys. Lett. B}\ }\textbf {\bibinfo {volume} {639}},\
  \bibinfo {pages} {151} (\bibinfo {year} {2006})},\ \bibinfo {note} {[Erratum:
  Phys.Lett.B 658, 285--289 (2008)]},\ \Eprint
  {https://arxiv.org/abs/hep-ex/0511054} {arXiv:hep-ex/0511054} \BibitemShut
  {NoStop}%
\bibitem [{\citenamefont {Aad}\ \emph {et~al.}(2016)\citenamefont {Aad} \emph
  {et~al.}}]{ATLAS:2016}%
  \BibitemOpen
  \bibfield  {author} {\bibinfo {author} {\bibfnamefont {G.}~\bibnamefont
  {Aad}} \emph {et~al.} (\bibinfo {collaboration} {ATLAS}),\ }\bibfield
  {title} {\bibinfo {title} {{Measurement of the inclusive isolated prompt
  photon cross section in pp collisions at $ \sqrt{s}=8 $ TeV with the ATLAS
  detector}},\ }\href {https://doi.org/10.1007/JHEP08(2016)005} {\bibfield
  {journal} {\bibinfo  {journal} {JHEP}\ }\textbf {\bibinfo {volume} {08}},\
  \bibinfo {pages} {005}}\BibitemShut {NoStop}%
\bibitem [{\citenamefont {Aad}\ \emph {et~al.}(2019)\citenamefont {Aad} \emph
  {et~al.}}]{ATLAS:2019}%
  \BibitemOpen
  \bibfield  {author} {\bibinfo {author} {\bibfnamefont {G.}~\bibnamefont
  {Aad}} \emph {et~al.} (\bibinfo {collaboration} {ATLAS}),\ }\bibfield
  {title} {\bibinfo {title} {{Measurement of the inclusive isolated-photon
  cross section in $pp$ collisions at $\sqrt{s}=13$ TeV using 36 fb$^{-1}$ of
  ATLAS data}},\ }\href {https://doi.org/10.1007/JHEP10(2019)203} {\bibfield
  {journal} {\bibinfo  {journal} {JHEP}\ }\textbf {\bibinfo {volume} {10}},\
  \bibinfo {pages} {203}},\ \Eprint {https://arxiv.org/abs/1908.02746}
  {arXiv:1908.02746 [hep-ex]} \BibitemShut {NoStop}%
\bibitem [{\citenamefont {Aad}\ \emph {et~al.}(2023)\citenamefont {Aad} \emph
  {et~al.}}]{ATLAS:2023}%
  \BibitemOpen
  \bibfield  {author} {\bibinfo {author} {\bibfnamefont {G.}~\bibnamefont
  {Aad}} \emph {et~al.} (\bibinfo {collaboration} {ATLAS}),\ }\bibfield
  {title} {\bibinfo {title} {{Inclusive-photon production and its dependence on
  photon isolation in $pp$ collisions at $\sqrt s=13$ TeV using 139 fb$^{-1}$
  of ATLAS data}},\ }\href {https://doi.org/10.1007/JHEP07(2023)086} {\bibfield
   {journal} {\bibinfo  {journal} {JHEP}\ }\textbf {\bibinfo {volume} {07}},\
  \bibinfo {pages} {086}},\ \Eprint {https://arxiv.org/abs/2302.00510}
  {arXiv:2302.00510 [hep-ex]} \BibitemShut {NoStop}%
\bibitem [{\citenamefont {Sirunyan}\ \emph {et~al.}(2019)\citenamefont
  {Sirunyan} \emph {et~al.}}]{CMS:2018}%
  \BibitemOpen
  \bibfield  {author} {\bibinfo {author} {\bibfnamefont {A.~M.}\ \bibnamefont
  {Sirunyan}} \emph {et~al.} (\bibinfo {collaboration} {CMS}),\ }\bibfield
  {title} {\bibinfo {title} {{Measurement of differential cross sections for
  inclusive isolated-photon and photon+jets production in proton-proton
  collisions at $\sqrt{s} =$ 13 TeV}},\ }\href
  {https://doi.org/10.1140/epjc/s10052-018-6482-9} {\bibfield  {journal}
  {\bibinfo  {journal} {Eur. Phys. J. C}\ }\textbf {\bibinfo {volume} {79}},\
  \bibinfo {pages} {20} (\bibinfo {year} {2019})},\ \Eprint
  {https://arxiv.org/abs/1807.00782} {arXiv:1807.00782 [hep-ex]} \BibitemShut
  {NoStop}%
\bibitem [{\citenamefont {Adare}\ \emph {et~al.}(2012)\citenamefont {Adare}
  \emph {et~al.}}]{PHENIX:2012}%
  \BibitemOpen
  \bibfield  {author} {\bibinfo {author} {\bibfnamefont {A.}~\bibnamefont
  {Adare}} \emph {et~al.} (\bibinfo {collaboration} {PHENIX}),\ }\bibfield
  {title} {\bibinfo {title} {{Direct-Photon Production in $p+p$ Collisions at
  $\sqrt{s}=200$ GeV at Midrapidity}},\ }\href
  {https://doi.org/10.1103/PhysRevD.86.072008} {\bibfield  {journal} {\bibinfo
  {journal} {Phys. Rev. D}\ }\textbf {\bibinfo {volume} {86}},\ \bibinfo
  {pages} {072008} (\bibinfo {year} {2012})},\ \Eprint
  {https://arxiv.org/abs/1205.5533} {arXiv:1205.5533 [hep-ex]} \BibitemShut
  {NoStop}%
\bibitem [{\citenamefont {Ballocchi}\ \emph {et~al.}(1998)\citenamefont
  {Ballocchi} \emph {et~al.}}]{UA6:1998}%
  \BibitemOpen
  \bibfield  {author} {\bibinfo {author} {\bibfnamefont {G.}~\bibnamefont
  {Ballocchi}} \emph {et~al.} (\bibinfo {collaboration} {UA6}),\ }\bibfield
  {title} {\bibinfo {title} {{Direct photon cross-sections in proton proton and
  anti-proton - proton interactions at S**(1/2) = 24.3-GeV}},\ }\href
  {https://doi.org/10.1016/S0370-2693(98)01001-6} {\bibfield  {journal}
  {\bibinfo  {journal} {Phys. Lett. B}\ }\textbf {\bibinfo {volume} {436}},\
  \bibinfo {pages} {222} (\bibinfo {year} {1998})}\BibitemShut {NoStop}%
\bibitem [{\citenamefont {Gordon}\ and\ \citenamefont
  {Vogelsang}(1994)}]{Gordon:1994}%
  \BibitemOpen
  \bibfield  {author} {\bibinfo {author} {\bibfnamefont {L.~E.}\ \bibnamefont
  {Gordon}}\ and\ \bibinfo {author} {\bibfnamefont {W.}~\bibnamefont
  {Vogelsang}},\ }\bibfield  {title} {\bibinfo {title} {{Polarized and
  unpolarized isolated prompt photon production beyond the leading order}},\
  }\href {https://doi.org/10.1103/PhysRevD.50.1901} {\bibfield  {journal}
  {\bibinfo  {journal} {Phys. Rev. D}\ }\textbf {\bibinfo {volume} {50}},\
  \bibinfo {pages} {1901} (\bibinfo {year} {1994})}\BibitemShut {NoStop}%
\bibitem [{\citenamefont {Catani}\ \emph {et~al.}(2002)\citenamefont {Catani},
  \citenamefont {Fontannaz}, \citenamefont {Guillet},\ and\ \citenamefont
  {Pilon}}]{Catani:2002}%
  \BibitemOpen
  \bibfield  {author} {\bibinfo {author} {\bibfnamefont {S.}~\bibnamefont
  {Catani}}, \bibinfo {author} {\bibfnamefont {M.}~\bibnamefont {Fontannaz}},
  \bibinfo {author} {\bibfnamefont {J.~P.}\ \bibnamefont {Guillet}},\ and\
  \bibinfo {author} {\bibfnamefont {E.}~\bibnamefont {Pilon}},\ }\bibfield
  {title} {\bibinfo {title} {{Cross-section of isolated prompt photons in
  hadron hadron collisions}},\ }\href
  {https://doi.org/10.1088/1126-6708/2002/05/028} {\bibfield  {journal}
  {\bibinfo  {journal} {JHEP}\ }\textbf {\bibinfo {volume} {05}},\ \bibinfo
  {pages} {028}}\BibitemShut {NoStop}%
\bibitem [{\citenamefont {Aurenche}\ \emph {et~al.}(2006)\citenamefont
  {Aurenche}, \citenamefont {Fontannaz}, \citenamefont {Guillet}, \citenamefont
  {Pilon},\ and\ \citenamefont {Werlen}}]{Aurenche:2006}%
  \BibitemOpen
  \bibfield  {author} {\bibinfo {author} {\bibfnamefont {P.}~\bibnamefont
  {Aurenche}}, \bibinfo {author} {\bibfnamefont {M.}~\bibnamefont {Fontannaz}},
  \bibinfo {author} {\bibfnamefont {J.-P.}\ \bibnamefont {Guillet}}, \bibinfo
  {author} {\bibfnamefont {E.}~\bibnamefont {Pilon}},\ and\ \bibinfo {author}
  {\bibfnamefont {M.}~\bibnamefont {Werlen}},\ }\bibfield  {title} {\bibinfo
  {title} {{A New critical study of photon production in hadronic
  collisions}},\ }\href {https://doi.org/10.1103/PhysRevD.73.094007} {\bibfield
   {journal} {\bibinfo  {journal} {Phys. Rev. D}\ }\textbf {\bibinfo {volume}
  {73}},\ \bibinfo {pages} {094007} (\bibinfo {year} {2006})}\BibitemShut
  {NoStop}%
\bibitem [{\citenamefont {Saleev}(2008)}]{Saleev:2008}%
  \BibitemOpen
  \bibfield  {author} {\bibinfo {author} {\bibfnamefont {V.~A.}\ \bibnamefont
  {Saleev}},\ }\bibfield  {title} {\bibinfo {title} {{Prompt photon
  photoproduction at HERA within the framework of the quark Reggeization
  hypothesis}},\ }\href {https://doi.org/10.1103/PhysRevD.78.114031} {\bibfield
   {journal} {\bibinfo  {journal} {Phys. Rev. D}\ }\textbf {\bibinfo {volume}
  {78}},\ \bibinfo {pages} {114031} (\bibinfo {year} {2008})}\BibitemShut
  {NoStop}%
\bibitem [{\citenamefont {Ganguli}\ \emph {et~al.}(2023)\citenamefont
  {Ganguli}, \citenamefont {van Hameren}, \citenamefont {Kotko},\ and\
  \citenamefont {Kutak}}]{Hameren:2023}%
  \BibitemOpen
  \bibfield  {author} {\bibinfo {author} {\bibfnamefont {I.}~\bibnamefont
  {Ganguli}}, \bibinfo {author} {\bibfnamefont {A.}~\bibnamefont {van
  Hameren}}, \bibinfo {author} {\bibfnamefont {P.}~\bibnamefont {Kotko}},\ and\
  \bibinfo {author} {\bibfnamefont {K.}~\bibnamefont {Kutak}},\ }\bibfield
  {title} {\bibinfo {title} {{Forward $\gamma $+jet production in proton-proton
  and proton-lead collisions at LHC within the FoCal calorimeter acceptance}},\
  }\href {https://doi.org/10.1140/epjc/s10052-023-12043-3} {\bibfield
  {journal} {\bibinfo  {journal} {Eur. Phys. J. C}\ }\textbf {\bibinfo {volume}
  {83}},\ \bibinfo {pages} {868} (\bibinfo {year} {2023})},\ \Eprint
  {https://arxiv.org/abs/2306.04706} {arXiv:2306.04706 [hep-ph]} \BibitemShut
  {NoStop}%
\bibitem [{\citenamefont {Arbuzov}\ \emph {et~al.}(2021)\citenamefont {Arbuzov}
  \emph {et~al.}}]{SPD}%
  \BibitemOpen
  \bibfield  {author} {\bibinfo {author} {\bibfnamefont {A.}~\bibnamefont
  {Arbuzov}} \emph {et~al.},\ }\bibfield  {title} {\bibinfo {title} {{On the
  physics potential to study the gluon content of proton and deuteron at NICA
  SPD}},\ }\href {https://doi.org/10.1016/j.ppnp.2021.103858} {\bibfield
  {journal} {\bibinfo  {journal} {Prog. Part. Nucl. Phys.}\ }\textbf {\bibinfo
  {volume} {119}},\ \bibinfo {pages} {103858} (\bibinfo {year}
  {2021})}\BibitemShut {NoStop}%
\bibitem [{\citenamefont {Gribov}\ and\ \citenamefont
  {Lipatov}(1972)}]{GL:1972}%
  \BibitemOpen
  \bibfield  {author} {\bibinfo {author} {\bibfnamefont {V.~N.}\ \bibnamefont
  {Gribov}}\ and\ \bibinfo {author} {\bibfnamefont {L.~N.}\ \bibnamefont
  {Lipatov}},\ }\bibfield  {title} {\bibinfo {title} {{Deep inelastic e p
  scattering in perturbation theory}},\ }\href@noop {} {\bibfield  {journal}
  {\bibinfo  {journal} {Sov. J. Nucl. Phys.}\ }\textbf {\bibinfo {volume}
  {15}},\ \bibinfo {pages} {438} (\bibinfo {year} {1972})}\BibitemShut
  {NoStop}%
\bibitem [{\citenamefont {Altarelli}\ and\ \citenamefont
  {Parisi}(1977)}]{AP:1977}%
  \BibitemOpen
  \bibfield  {author} {\bibinfo {author} {\bibfnamefont {G.}~\bibnamefont
  {Altarelli}}\ and\ \bibinfo {author} {\bibfnamefont {G.}~\bibnamefont
  {Parisi}},\ }\bibfield  {title} {\bibinfo {title} {{Asymptotic Freedom in
  Parton Language}},\ }\href {https://doi.org/10.1016/0550-3213(77)90384-4}
  {\bibfield  {journal} {\bibinfo  {journal} {Nucl. Phys. B}\ }\textbf
  {\bibinfo {volume} {126}},\ \bibinfo {pages} {298} (\bibinfo {year}
  {1977})}\BibitemShut {NoStop}%
\bibitem [{\citenamefont {Dokshitzer}(1977)}]{D:1977}%
  \BibitemOpen
  \bibfield  {author} {\bibinfo {author} {\bibfnamefont {Y.~L.}\ \bibnamefont
  {Dokshitzer}},\ }\bibfield  {title} {\bibinfo {title} {{Calculation of the
  Structure Functions for Deep Inelastic Scattering and e+ e- Annihilation by
  Perturbation Theory in Quantum Chromodynamics.}},\ }\href@noop {} {\bibfield
  {journal} {\bibinfo  {journal} {Sov. Phys. JETP}\ }\textbf {\bibinfo {volume}
  {46}},\ \bibinfo {pages} {641} (\bibinfo {year} {1977})}\BibitemShut
  {NoStop}%
\bibitem [{\citenamefont {Gribov}\ \emph {et~al.}(1983)\citenamefont {Gribov},
  \citenamefont {Levin},\ and\ \citenamefont {Ryskin}}]{HEF1}%
  \BibitemOpen
  \bibfield  {author} {\bibinfo {author} {\bibfnamefont {L.~V.}\ \bibnamefont
  {Gribov}}, \bibinfo {author} {\bibfnamefont {E.~M.}\ \bibnamefont {Levin}},\
  and\ \bibinfo {author} {\bibfnamefont {M.~G.}\ \bibnamefont {Ryskin}},\
  }\bibfield  {title} {\bibinfo {title} {{Semihard Processes in QCD}},\ }\href
  {https://doi.org/10.1016/0370-1573(83)90022-4} {\bibfield  {journal}
  {\bibinfo  {journal} {Phys. Rept.}\ }\textbf {\bibinfo {volume} {100}},\
  \bibinfo {pages} {1} (\bibinfo {year} {1983})}\BibitemShut {NoStop}%
\bibitem [{\citenamefont {Collins}\ and\ \citenamefont {Ellis}(1991)}]{HEF2}%
  \BibitemOpen
  \bibfield  {author} {\bibinfo {author} {\bibfnamefont {J.~C.}\ \bibnamefont
  {Collins}}\ and\ \bibinfo {author} {\bibfnamefont {R.~K.}\ \bibnamefont
  {Ellis}},\ }\bibfield  {title} {\bibinfo {title} {Heavy quark production in
  very high-energy hadron collisions},\ }\href
  {https://doi.org/10.1016/0550-3213(91)90288-9} {\bibfield  {journal}
  {\bibinfo  {journal} {Nucl. Phys. B}\ }\textbf {\bibinfo {volume} {360}},\
  \bibinfo {pages} {3} (\bibinfo {year} {1991})}\BibitemShut {NoStop}%
\bibitem [{\citenamefont {Catani}\ and\ \citenamefont {Hautmann}(1994)}]{HEF3}%
  \BibitemOpen
  \bibfield  {author} {\bibinfo {author} {\bibfnamefont {S.}~\bibnamefont
  {Catani}}\ and\ \bibinfo {author} {\bibfnamefont {F.}~\bibnamefont
  {Hautmann}},\ }\bibfield  {title} {\bibinfo {title} {{High-energy
  factorization and small x deep inelastic scattering beyond leading order}},\
  }\href {https://doi.org/10.1016/0550-3213(94)90636-X} {\bibfield  {journal}
  {\bibinfo  {journal} {Nucl. Phys. B}\ }\textbf {\bibinfo {volume} {427}},\
  \bibinfo {pages} {475} (\bibinfo {year} {1994})},\ \Eprint
  {https://arxiv.org/abs/hep-ph/9405388} {arXiv:hep-ph/9405388} \BibitemShut
  {NoStop}%
\bibitem [{\citenamefont {Lipatov}(1976)}]{BFKL1}%
  \BibitemOpen
  \bibfield  {author} {\bibinfo {author} {\bibfnamefont {L.~N.}\ \bibnamefont
  {Lipatov}},\ }\bibfield  {title} {\bibinfo {title} {{Reggeization of the
  Vector Meson and the Vacuum Singularity in Nonabelian Gauge Theories}},\
  }\href@noop {} {\bibfield  {journal} {\bibinfo  {journal} {Sov. J. Nucl.
  Phys.}\ }\textbf {\bibinfo {volume} {23}},\ \bibinfo {pages} {338} (\bibinfo
  {year} {1976})}\BibitemShut {NoStop}%
\bibitem [{\citenamefont {Kuraev}\ \emph {et~al.}(1976)\citenamefont {Kuraev},
  \citenamefont {Lipatov},\ and\ \citenamefont {Fadin}}]{BFKL2}%
  \BibitemOpen
  \bibfield  {author} {\bibinfo {author} {\bibfnamefont {E.~A.}\ \bibnamefont
  {Kuraev}}, \bibinfo {author} {\bibfnamefont {L.~N.}\ \bibnamefont
  {Lipatov}},\ and\ \bibinfo {author} {\bibfnamefont {V.~S.}\ \bibnamefont
  {Fadin}},\ }\bibfield  {title} {\bibinfo {title} {{Multi - Reggeon Processes
  in the Yang-Mills Theory}},\ }\href@noop {} {\bibfield  {journal} {\bibinfo
  {journal} {Sov. Phys. JETP}\ }\textbf {\bibinfo {volume} {44}},\ \bibinfo
  {pages} {443} (\bibinfo {year} {1976})}\BibitemShut {NoStop}%
\bibitem [{\citenamefont {Kuraev}\ \emph {et~al.}(1977)\citenamefont {Kuraev},
  \citenamefont {Lipatov},\ and\ \citenamefont {Fadin}}]{BFKL3}%
  \BibitemOpen
  \bibfield  {author} {\bibinfo {author} {\bibfnamefont {E.~A.}\ \bibnamefont
  {Kuraev}}, \bibinfo {author} {\bibfnamefont {L.~N.}\ \bibnamefont
  {Lipatov}},\ and\ \bibinfo {author} {\bibfnamefont {V.~S.}\ \bibnamefont
  {Fadin}},\ }\bibfield  {title} {\bibinfo {title} {{The Pomeranchuk
  Singularity in Nonabelian Gauge Theories}},\ }\href@noop {} {\bibfield
  {journal} {\bibinfo  {journal} {Sov. Phys. JETP}\ }\textbf {\bibinfo {volume}
  {45}},\ \bibinfo {pages} {199} (\bibinfo {year} {1977})}\BibitemShut
  {NoStop}%
\bibitem [{\citenamefont {Balitsky}\ and\ \citenamefont
  {Lipatov}(1978)}]{BFKL4}%
  \BibitemOpen
  \bibfield  {author} {\bibinfo {author} {\bibfnamefont {I.~I.}\ \bibnamefont
  {Balitsky}}\ and\ \bibinfo {author} {\bibfnamefont {L.~N.}\ \bibnamefont
  {Lipatov}},\ }\bibfield  {title} {\bibinfo {title} {{The Pomeranchuk
  Singularity in Quantum Chromodynamics}},\ }\href@noop {} {\bibfield
  {journal} {\bibinfo  {journal} {Sov. J. Nucl. Phys.}\ }\textbf {\bibinfo
  {volume} {28}},\ \bibinfo {pages} {822} (\bibinfo {year} {1978})}\BibitemShut
  {NoStop}%
\bibitem [{\citenamefont {Kniehl}\ \emph {et~al.}(2006)\citenamefont {Kniehl},
  \citenamefont {Vasin},\ and\ \citenamefont {Saleev}}]{PRA1}%
  \BibitemOpen
  \bibfield  {author} {\bibinfo {author} {\bibfnamefont {B.~A.}\ \bibnamefont
  {Kniehl}}, \bibinfo {author} {\bibfnamefont {D.~V.}\ \bibnamefont {Vasin}},\
  and\ \bibinfo {author} {\bibfnamefont {V.~A.}\ \bibnamefont {Saleev}},\
  }\bibfield  {title} {\bibinfo {title} {{Charmonium production at high energy
  in the $k_{T}$ -factorization approach}},\ }\href
  {https://doi.org/10.1103/PhysRevD.73.074022} {\bibfield  {journal} {\bibinfo
  {journal} {Phys. Rev. D}\ }\textbf {\bibinfo {volume} {73}},\ \bibinfo
  {pages} {074022} (\bibinfo {year} {2006})},\ \Eprint
  {https://arxiv.org/abs/hep-ph/0602179} {arXiv:hep-ph/0602179} \BibitemShut
  {NoStop}%
\bibitem [{\citenamefont {Nefedov}\ \emph {et~al.}(2013)\citenamefont
  {Nefedov}, \citenamefont {Saleev},\ and\ \citenamefont {Shipilova}}]{PRA2}%
  \BibitemOpen
  \bibfield  {author} {\bibinfo {author} {\bibfnamefont {M.~A.}\ \bibnamefont
  {Nefedov}}, \bibinfo {author} {\bibfnamefont {V.~A.}\ \bibnamefont
  {Saleev}},\ and\ \bibinfo {author} {\bibfnamefont {A.~V.}\ \bibnamefont
  {Shipilova}},\ }\bibfield  {title} {\bibinfo {title} {{Dijet azimuthal
  decorrelations at the LHC in the parton Reggeization approach}},\ }\href
  {https://doi.org/10.1103/PhysRevD.87.094030} {\bibfield  {journal} {\bibinfo
  {journal} {Phys. Rev. D}\ }\textbf {\bibinfo {volume} {87}},\ \bibinfo
  {pages} {094030} (\bibinfo {year} {2013})}\BibitemShut {NoStop}%
\bibitem [{\citenamefont {Karpishkov}\ \emph {et~al.}(2017)\citenamefont
  {Karpishkov}, \citenamefont {Nefedov},\ and\ \citenamefont {Saleev}}]{PRA3}%
  \BibitemOpen
  \bibfield  {author} {\bibinfo {author} {\bibfnamefont {A.~V.}\ \bibnamefont
  {Karpishkov}}, \bibinfo {author} {\bibfnamefont {M.~A.}\ \bibnamefont
  {Nefedov}},\ and\ \bibinfo {author} {\bibfnamefont {V.~A.}\ \bibnamefont
  {Saleev}},\ }\bibfield  {title} {\bibinfo {title} {{$B{\bar B}$ angular
  correlations at the LHC in parton Reggeization approach merged with
  higher-order matrix elements}},\ }\href
  {https://doi.org/10.1103/PhysRevD.96.096019} {\bibfield  {journal} {\bibinfo
  {journal} {Phys. Rev. D}\ }\textbf {\bibinfo {volume} {96}},\ \bibinfo
  {pages} {096019} (\bibinfo {year} {2017})},\ \Eprint
  {https://arxiv.org/abs/1707.04068} {arXiv:1707.04068 [hep-ph]} \BibitemShut
  {NoStop}%
\bibitem [{\citenamefont {Nefedov}\ and\ \citenamefont
  {Saleev}(2020)}]{NS:2020}%
  \BibitemOpen
  \bibfield  {author} {\bibinfo {author} {\bibfnamefont {M.~A.}\ \bibnamefont
  {Nefedov}}\ and\ \bibinfo {author} {\bibfnamefont {V.~A.}\ \bibnamefont
  {Saleev}},\ }\bibfield  {title} {\bibinfo {title} {{High--Energy
  Factorization for Drell--Yan process in $pp$ and $p{\bar p}$ collisions with
  new Unintegrated PDFs}},\ }\href
  {https://doi.org/10.1103/PhysRevD.102.114018} {\bibfield  {journal} {\bibinfo
   {journal} {Phys. Rev. D}\ }\textbf {\bibinfo {volume} {102}},\ \bibinfo
  {pages} {114018} (\bibinfo {year} {2020})}\BibitemShut {NoStop}%
\bibitem [{\citenamefont {Nefedov}\ and\ \citenamefont
  {Saleev}(2015)}]{NS:2016}%
  \BibitemOpen
  \bibfield  {author} {\bibinfo {author} {\bibfnamefont {M.}~\bibnamefont
  {Nefedov}}\ and\ \bibinfo {author} {\bibfnamefont {V.}~\bibnamefont
  {Saleev}},\ }\bibfield  {title} {\bibinfo {title} {{Diphoton production at
  the Tevatron and the LHC in the NLO approximation of the parton Reggeization
  approach}},\ }\href {https://doi.org/10.1103/PhysRevD.92.094033} {\bibfield
  {journal} {\bibinfo  {journal} {Phys. Rev. D}\ }\textbf {\bibinfo {volume}
  {92}},\ \bibinfo {pages} {094033} (\bibinfo {year} {2015})}\BibitemShut
  {NoStop}%
\bibitem [{\citenamefont {Lipatov}(1995)}]{Lipatov:1}%
  \BibitemOpen
  \bibfield  {author} {\bibinfo {author} {\bibfnamefont {L.~N.}\ \bibnamefont
  {Lipatov}},\ }\bibfield  {title} {\bibinfo {title} {{Gauge invariant
  effective action for high--energy processes in QCD}},\ }\href
  {https://doi.org/10.1016/0550-3213(95)00390-E} {\bibfield  {journal}
  {\bibinfo  {journal} {Nucl. Phys. B}\ }\textbf {\bibinfo {volume} {452}},\
  \bibinfo {pages} {369} (\bibinfo {year} {1995})}\BibitemShut {NoStop}%
\bibitem [{\citenamefont {Lipatov}\ and\ \citenamefont
  {Vyazovsky}(2001)}]{Lipatov:2}%
  \BibitemOpen
  \bibfield  {author} {\bibinfo {author} {\bibfnamefont {L.~N.}\ \bibnamefont
  {Lipatov}}\ and\ \bibinfo {author} {\bibfnamefont {M.~I.}\ \bibnamefont
  {Vyazovsky}},\ }\bibfield  {title} {\bibinfo {title} {{QuasimultiRegge
  processes with a quark exchange in the $t$ channel}},\ }\href
  {https://doi.org/10.1016/S0550-3213(00)00709-4} {\bibfield  {journal}
  {\bibinfo  {journal} {Nucl. Phys. B}\ }\textbf {\bibinfo {volume} {597}},\
  \bibinfo {pages} {399} (\bibinfo {year} {2001})}\BibitemShut {NoStop}%
\bibitem [{\citenamefont {Antonov}\ \emph {et~al.}(2005)\citenamefont
  {Antonov}, \citenamefont {Lipatov}, \citenamefont {Kuraev},\ and\
  \citenamefont {Cherednikov}}]{Lipatov:3}%
  \BibitemOpen
  \bibfield  {author} {\bibinfo {author} {\bibfnamefont {E.~N.}\ \bibnamefont
  {Antonov}}, \bibinfo {author} {\bibfnamefont {L.~N.}\ \bibnamefont
  {Lipatov}}, \bibinfo {author} {\bibfnamefont {E.~A.}\ \bibnamefont
  {Kuraev}},\ and\ \bibinfo {author} {\bibfnamefont {I.~O.}\ \bibnamefont
  {Cherednikov}},\ }\bibfield  {title} {\bibinfo {title} {{Feynman rules for
  effective Regge action}},\ }\href
  {https://doi.org/10.1016/j.nuclphysb.2005.013} {\bibfield  {journal}
  {\bibinfo  {journal} {Nucl. Phys. B}\ }\textbf {\bibinfo {volume} {721}},\
  \bibinfo {pages} {111} (\bibinfo {year} {2005})}\BibitemShut {NoStop}%
\bibitem [{\citenamefont {Kimber}\ \emph {et~al.}(2001)\citenamefont {Kimber},
  \citenamefont {Martin},\ and\ \citenamefont {Ryskin}}]{KMR}%
  \BibitemOpen
  \bibfield  {author} {\bibinfo {author} {\bibfnamefont {M.~A.}\ \bibnamefont
  {Kimber}}, \bibinfo {author} {\bibfnamefont {A.~D.}\ \bibnamefont {Martin}},\
  and\ \bibinfo {author} {\bibfnamefont {M.~G.}\ \bibnamefont {Ryskin}},\
  }\bibfield  {title} {\bibinfo {title} {{Unintegrated parton distributions}},\
  }\href {https://doi.org/10.1103/PhysRevD.63.114027} {\bibfield  {journal}
  {\bibinfo  {journal} {Phys. Rev. D}\ }\textbf {\bibinfo {volume} {63}},\
  \bibinfo {pages} {114027} (\bibinfo {year} {2001})}\BibitemShut {NoStop}%
\bibitem [{\citenamefont {Watt}\ \emph {et~al.}(2003)\citenamefont {Watt},
  \citenamefont {Martin},\ and\ \citenamefont {Ryskin}}]{MRW}%
  \BibitemOpen
  \bibfield  {author} {\bibinfo {author} {\bibfnamefont {G.}~\bibnamefont
  {Watt}}, \bibinfo {author} {\bibfnamefont {A.~D.}\ \bibnamefont {Martin}},\
  and\ \bibinfo {author} {\bibfnamefont {M.~G.}\ \bibnamefont {Ryskin}},\
  }\bibfield  {title} {\bibinfo {title} {{Unintegrated parton distributions and
  inclusive jet production at HERA}},\ }\href
  {https://doi.org/10.1140/epjc/s2003-01320-4} {\bibfield  {journal} {\bibinfo
  {journal} {Eur. Phys. J. C}\ }\textbf {\bibinfo {volume} {31}},\ \bibinfo
  {pages} {73} (\bibinfo {year} {2003})}\BibitemShut {NoStop}%
\bibitem [{\citenamefont {Collins}\ \emph {et~al.}(1985)\citenamefont
  {Collins}, \citenamefont {Soper},\ and\ \citenamefont {Sterman}}]{CSS}%
  \BibitemOpen
  \bibfield  {author} {\bibinfo {author} {\bibfnamefont {J.~C.}\ \bibnamefont
  {Collins}}, \bibinfo {author} {\bibfnamefont {D.~E.}\ \bibnamefont {Soper}},\
  and\ \bibinfo {author} {\bibfnamefont {G.~F.}\ \bibnamefont {Sterman}},\
  }\bibfield  {title} {\bibinfo {title} {{Transverse Momentum Distribution in
  Drell-Yan Pair and W and Z Boson Production}},\ }\href
  {https://doi.org/10.1016/0550-3213(85)90479-1} {\bibfield  {journal}
  {\bibinfo  {journal} {Nucl. Phys. B}\ }\textbf {\bibinfo {volume} {250}},\
  \bibinfo {pages} {199} (\bibinfo {year} {1985})}\BibitemShut {NoStop}%
\bibitem [{\citenamefont {Collins}(2023)}]{Collins:2011}%
  \BibitemOpen
  \bibfield  {author} {\bibinfo {author} {\bibfnamefont {J.}~\bibnamefont
  {Collins}},\ }\href {https://doi.org/10.1017/9781009401845} {}\bibinfo
  {series} {Cambridge Monographs on Particle Physics, Nuclear Physics and
  Cosmology}, Vol.~\bibinfo {volume} {32}\ (\bibinfo  {publisher} {Cambridge
  University Press},\ \bibinfo {year} {2023})\BibitemShut {NoStop}%
\bibitem [{\citenamefont {Kniehl}\ \emph {et~al.}(2014)\citenamefont {Kniehl},
  \citenamefont {Nefedov},\ and\ \citenamefont {Saleev}}]{NS:2014}%
  \BibitemOpen
  \bibfield  {author} {\bibinfo {author} {\bibfnamefont {B.~A.}\ \bibnamefont
  {Kniehl}}, \bibinfo {author} {\bibfnamefont {M.~A.}\ \bibnamefont
  {Nefedov}},\ and\ \bibinfo {author} {\bibfnamefont {V.~A.}\ \bibnamefont
  {Saleev}},\ }\bibfield  {title} {\bibinfo {title} {{Prompt-photon plus jet
  associated photoproduction at HERA in the parton Reggeization approach}},\
  }\href {https://doi.org/10.1103/PhysRevD.89.114016} {\bibfield  {journal}
  {\bibinfo  {journal} {Phys. Rev. D}\ }\textbf {\bibinfo {volume} {89}},\
  \bibinfo {pages} {114016} (\bibinfo {year} {2014})},\ \Eprint
  {https://arxiv.org/abs/1404.3513} {arXiv:1404.3513 [hep-ph]} \BibitemShut
  {NoStop}%
\bibitem [{\citenamefont {Saleev}(2009)}]{Saleev:2009}%
  \BibitemOpen
  \bibfield  {author} {\bibinfo {author} {\bibfnamefont {V.~A.}\ \bibnamefont
  {Saleev}},\ }\bibfield  {title} {\bibinfo {title} {{Diphoton production at
  Tevatron in the quasi-multi-Regge-kinematics approach}},\ }\href
  {https://doi.org/10.1103/PhysRevD.80.114016} {\bibfield  {journal} {\bibinfo
  {journal} {Phys. Rev. D}\ }\textbf {\bibinfo {volume} {80}},\ \bibinfo
  {pages} {114016} (\bibinfo {year} {2009})}\BibitemShut {NoStop}%
\bibitem [{\citenamefont {van Hameren}(2018)}]{katie}%
  \BibitemOpen
  \bibfield  {author} {\bibinfo {author} {\bibfnamefont {A.}~\bibnamefont {van
  Hameren}},\ }\bibfield  {title} {\bibinfo {title} {{KaTie : For parton-level
  event generation with $k_T$-dependent initial states}},\ }\href
  {https://doi.org/10.1016/j.cpc.2017.11.005} {\bibfield  {journal} {\bibinfo
  {journal} {Comput. Phys. Commun.}\ }\textbf {\bibinfo {volume} {224}},\
  \bibinfo {pages} {371} (\bibinfo {year} {2018})},\ \Eprint
  {https://arxiv.org/abs/1611.00680} {arXiv:1611.00680 [hep-ph]} \BibitemShut
  {NoStop}%
\bibitem [{\citenamefont {Karpishkov}\ and\ \citenamefont
  {Saleev}(2022)}]{KS:2022}%
  \BibitemOpen
  \bibfield  {author} {\bibinfo {author} {\bibfnamefont {A.}~\bibnamefont
  {Karpishkov}}\ and\ \bibinfo {author} {\bibfnamefont {V.}~\bibnamefont
  {Saleev}},\ }\bibfield  {title} {\bibinfo {title} {{Production of three
  isolated photons in the parton Reggeization approach at high energies}},\
  }\href {https://doi.org/10.1103/PhysRevD.106.054036} {\bibfield  {journal}
  {\bibinfo  {journal} {Phys. Rev. D}\ }\textbf {\bibinfo {volume} {106}},\
  \bibinfo {pages} {054036} (\bibinfo {year} {2022})},\ \Eprint
  {https://arxiv.org/abs/2205.12773} {arXiv:2205.12773 [hep-ph]} \BibitemShut
  {NoStop}%
\bibitem [{\citenamefont {Lipatov}(1997)}]{Lipatov:1996}%
  \BibitemOpen
  \bibfield  {author} {\bibinfo {author} {\bibfnamefont {L.~N.}\ \bibnamefont
  {Lipatov}},\ }\bibfield  {title} {\bibinfo {title} {{Small x physics in
  perturbative QCD}},\ }\href {https://doi.org/10.1016/S0370-1573(96)00045-2}
  {\bibfield  {journal} {\bibinfo  {journal} {Phys. Rept.}\ }\textbf {\bibinfo
  {volume} {286}},\ \bibinfo {pages} {131} (\bibinfo {year} {1997})},\ \Eprint
  {https://arxiv.org/abs/hep-ph/9610276} {arXiv:hep-ph/9610276} \BibitemShut
  {NoStop}%
\bibitem [{\citenamefont {Hahn}(2001)}]{FeynArts}%
  \BibitemOpen
  \bibfield  {author} {\bibinfo {author} {\bibfnamefont {T.}~\bibnamefont
  {Hahn}},\ }\bibfield  {title} {\bibinfo {title} {{Generating Feynman diagrams
  and amplitudes with FeynArts 3}},\ }\href
  {https://doi.org/10.1016/S0010-4655(01)00290-9} {\bibfield  {journal}
  {\bibinfo  {journal} {Comput. Phys. Commun.}\ }\textbf {\bibinfo {volume}
  {140}},\ \bibinfo {pages} {418} (\bibinfo {year} {2001})}\BibitemShut
  {NoStop}%
\bibitem [{\citenamefont {Fadin}\ and\ \citenamefont
  {Sherman}(1977)}]{Fadin:1977}%
  \BibitemOpen
  \bibfield  {author} {\bibinfo {author} {\bibfnamefont {V.~S.}\ \bibnamefont
  {Fadin}}\ and\ \bibinfo {author} {\bibfnamefont {V.~E.}\ \bibnamefont
  {Sherman}},\ }\bibfield  {title} {\bibinfo {title} {{Processes Involving
  Fermion Exchange in Nonabelian Gauge Theories}},\ }\href@noop {} {\bibfield
  {journal} {\bibinfo  {journal} {Zh. Eksp. Teor. Fiz.}\ }\textbf {\bibinfo
  {volume} {72}},\ \bibinfo {pages} {1640} (\bibinfo {year}
  {1977})}\BibitemShut {NoStop}%
\bibitem [{\citenamefont {Nefedov}\ and\ \citenamefont
  {Saleev}(2017)}]{NS:2017}%
  \BibitemOpen
  \bibfield  {author} {\bibinfo {author} {\bibfnamefont {M.}~\bibnamefont
  {Nefedov}}\ and\ \bibinfo {author} {\bibfnamefont {V.}~\bibnamefont
  {Saleev}},\ }\bibfield  {title} {\bibinfo {title} {{On the one-loop
  calculations with Reggeized quarks}},\ }\href
  {https://doi.org/10.1142/S0217732317502078} {\bibfield  {journal} {\bibinfo
  {journal} {Mod. Phys. Lett. A}\ }\textbf {\bibinfo {volume} {32}},\ \bibinfo
  {pages} {1750207} (\bibinfo {year} {2017})}\BibitemShut {NoStop}%
\bibitem [{\citenamefont {Nefedov}\ and\ \citenamefont
  {Saleev}(2018)}]{NS:2018}%
  \BibitemOpen
  \bibfield  {author} {\bibinfo {author} {\bibfnamefont {M.}~\bibnamefont
  {Nefedov}}\ and\ \bibinfo {author} {\bibfnamefont {V.}~\bibnamefont
  {Saleev}},\ }\bibfield  {title} {\bibinfo {title} {{From LO to NLO in the
  parton Reggeization approach}},\ }\href
  {https://doi.org/10.1051/epjconf/201819104007} {\bibfield  {journal}
  {\bibinfo  {journal} {EPJ Web Conf.}\ }\textbf {\bibinfo {volume} {191}},\
  \bibinfo {pages} {04007} (\bibinfo {year} {2018})}\BibitemShut {NoStop}%
\bibitem [{\citenamefont {Nefedov}(2020)}]{MN:2020}%
  \BibitemOpen
  \bibfield  {author} {\bibinfo {author} {\bibfnamefont {M.~A.}\ \bibnamefont
  {Nefedov}},\ }\bibfield  {title} {\bibinfo {title} {{Towards stability of NLO
  corrections in High-Energy Factorization via Modified Multi-Regge Kinematics
  approximation}},\ }\href {https://doi.org/10.1007/JHEP08(2020)055} {\bibfield
   {journal} {\bibinfo  {journal} {JHEP}\ }\textbf {\bibinfo {volume} {08}},\
  \bibinfo {pages} {055}},\ \Eprint {https://arxiv.org/abs/2003.02194}
  {arXiv:2003.02194 [hep-ph]} \BibitemShut {NoStop}%
\bibitem [{\citenamefont {Nefedov}(2021)}]{MN:2021}%
  \BibitemOpen
  \bibfield  {author} {\bibinfo {author} {\bibfnamefont {M.}~\bibnamefont
  {Nefedov}},\ }\bibfield  {title} {\bibinfo {title} {{Sudakov resummation from
  the BFKL evolution}},\ }\href {https://doi.org/10.1103/PhysRevD.104.054039}
  {\bibfield  {journal} {\bibinfo  {journal} {Phys. Rev. D}\ }\textbf {\bibinfo
  {volume} {104}},\ \bibinfo {pages} {054039} (\bibinfo {year} {2021})},\
  \Eprint {https://arxiv.org/abs/2105.13915} {arXiv:2105.13915 [hep-ph]}
  \BibitemShut {NoStop}%
\bibitem [{\citenamefont {Saleev}\ and\ \citenamefont
  {Shipilova}(2011)}]{Saleev:2011}%
  \BibitemOpen
  \bibfield  {author} {\bibinfo {author} {\bibfnamefont {V.~A.}\ \bibnamefont
  {Saleev}}\ and\ \bibinfo {author} {\bibfnamefont {A.~V.}\ \bibnamefont
  {Shipilova}},\ }\bibfield  {title} {\bibinfo {title} {{Production of b-quark
  jets at the Tevatron collider in the Regge limit of QCD}},\ }\href
  {https://doi.org/10.1134/S1063778810121038} {\bibfield  {journal} {\bibinfo
  {journal} {Phys. Atom. Nucl.}\ }\textbf {\bibinfo {volume} {74}},\ \bibinfo
  {pages} {151} (\bibinfo {year} {2011})}\BibitemShut {NoStop}%
\bibitem [{\citenamefont {Ioffe}\ \emph {et~al.}(2010)\citenamefont {Ioffe},
  \citenamefont {Fadin},\ and\ \citenamefont {Lipatov}}]{Lipatov:10}%
  \BibitemOpen
  \bibfield  {author} {\bibinfo {author} {\bibfnamefont {B.~L.}\ \bibnamefont
  {Ioffe}}, \bibinfo {author} {\bibfnamefont {V.~S.}\ \bibnamefont {Fadin}},\
  and\ \bibinfo {author} {\bibfnamefont {L.~N.}\ \bibnamefont {Lipatov}},\
  }\href {https://doi.org/10.1017/CBO9780511711817} {}\ (\bibinfo  {publisher}
  {Cambridge Univ. Press},\ \bibinfo {year} {2010})\BibitemShut {NoStop}%
\bibitem [{\citenamefont {Kovchegov}\ and\ \citenamefont
  {Levin}(2013)}]{Levin:12}%
  \BibitemOpen
  \bibfield  {author} {\bibinfo {author} {\bibfnamefont {Y.~V.}\ \bibnamefont
  {Kovchegov}}\ and\ \bibinfo {author} {\bibfnamefont {E.}~\bibnamefont
  {Levin}},\ }\href {https://doi.org/10.1017/9781009291446} {}Vol.~\bibinfo
  {volume} {33}\ (\bibinfo  {publisher} {Oxford University Press},\ \bibinfo
  {year} {2013})\BibitemShut {NoStop}%
\bibitem [{\citenamefont {Mueller}\ and\ \citenamefont
  {Navelet}(1987)}]{Mueller:1986}%
  \BibitemOpen
  \bibfield  {author} {\bibinfo {author} {\bibfnamefont {A.~H.}\ \bibnamefont
  {Mueller}}\ and\ \bibinfo {author} {\bibfnamefont {H.}~\bibnamefont
  {Navelet}},\ }\bibfield  {title} {\bibinfo {title} {{An Inclusive Minijet
  Cross-Section and the Bare Pomeron in QCD}},\ }\href
  {https://doi.org/10.1016/0550-3213(87)90705-X} {\bibfield  {journal}
  {\bibinfo  {journal} {Nucl. Phys. B}\ }\textbf {\bibinfo {volume} {282}},\
  \bibinfo {pages} {727} (\bibinfo {year} {1987})}\BibitemShut {NoStop}%
\bibitem [{\citenamefont {Hahn}(2005)}]{Cuba}%
  \BibitemOpen
  \bibfield  {author} {\bibinfo {author} {\bibfnamefont {T.}~\bibnamefont
  {Hahn}},\ }\bibfield  {title} {\bibinfo {title} {{CUBA: A Library for
  multidimensional numerical integration}},\ }\href
  {https://doi.org/10.1016/j.cpc.2005.01.010} {\bibfield  {journal} {\bibinfo
  {journal} {Comput. Phys. Commun.}\ }\textbf {\bibinfo {volume} {168}},\
  \bibinfo {pages} {78} (\bibinfo {year} {2005})}\BibitemShut {NoStop}%
\end{thebibliography}%

\newpage

\end{document}